\documentclass[12pt]{article}
\pdfoutput=1 
\usepackage{amsmath}
\usepackage{amssymb}
\usepackage{graphicx}
\usepackage{enumerate}
\usepackage{slashed, bm}
\usepackage{hyperref, xcolor}
\definecolor{dark}{rgb}{0.10,0.2,0.3}
\definecolor{magenta}{rgb}{0.7,0.1,0.3}
\definecolor{purpure}{rgb}{0.5,0.15,0.3}
\usepackage[font=small,format=plain,labelfont=bf,up,textfont=it,up]{caption}
\usepackage{hyperref, cite}
\hypersetup{colorlinks,
citecolor=blue,
filecolor=blue,
linkcolor=magenta,
urlcolor=purpure,hyperfootnotes=true,pdftex}

\usepackage[normalem]{ulem}
\usepackage{subfig}
         \let\g = \gamma     \let\e = \epsilon

\newcommand{\nn}{\nonumber}
\newcommand{\beq}{\begin{equation}}
\newcommand{\eeq}{\end{equation}}
\newcommand{\bea}{\begin{eqnarray}}
\newcommand{\eea}{\end{eqnarray}}

\newcommand{\AL}[1]{\langle#1|}
\newcommand{\AR}[1]{|#1\rangle}
\newcommand{\SL}[1]{[#1|}
\newcommand{\SR}[1]{|#1]}
\renewcommand{\AA}[1]{\langle#1\rangle}
\newcommand{\SSS}[1]{[#1]}
\newcommand{\AS}[1]{\langle#1]}

\renewcommand{\sp}[2]{\,#1\!\cdot\!#2\,}
\newcommand{\slashk}{k \! \! \!  /}
\newcommand{\slashl}{l \! \! \!  /}
\newcommand{\slashK}{K \! \! \!  /}
\newcommand{\slashp}{p \! \! \!  /}
\newcommand{\slashq}{q \! \! \!  /}

\newcommand{\slashn}{n \! \! \!  /}
\newcommand{\slasheps}{\epsilon \! \! \!  /}
\newcommand{\kapp}{\kappa}

\newcommand{\kstr}{\kappa^*}

\newcommand{\qb}{\bar{q}}

\newcommand{\Amp}{\mathcal{A}}

\newcommand{\Pbb}{\mathbb{P}} 
\newcommand{\qt}{{ \bm q}} 
\newcommand{\pt}{{ \bm p}}
\newcommand{\ptt}{{\tilde{\bm p}}}
\newcommand{\qtt}{{\tilde{\bm q}}} 
\newcommand{\kt}{{\bm k}} %

\title{
\vspace{-4.0cm}
\begin{flushright}
{\small IFJPAN-IV-2017-25$\qquad\quad$}\\
\end{flushright}
\vspace{1.5cm}
\boldmath \large \bf TMD splitting functions in $k_T$ factorization: \\  the real
contribution to  the gluon-to-gluon splitting}

\author{M.~Hentschinski${}^a$, A.~Kusina${}^{b}$, K.~Kutak${}^b$, M.~Serino${}^{b,c}$
\bigskip \\
${}^a$Departamento de Actuaria, F\'isica y Matem\'aticas, \\
Universidad de las Americas Puebla, \\ Santa Catarina Martir, 72820
Puebla, Mexico\\
${}^b$ The H. Niewodnicza\'nski Institute of Nuclear Physics, \\
Polish Academy of Sciences, \\ ul. Radzikowskiego 152, 31-342, Cracow,
Poland\\
${}^c$ Department of Physics, \\  Ben Gurion University of the
Negev, \\ Beer Sheva 8410501, Israel
}

\begin{document}

\maketitle \flushbottom
\begin{abstract}
  {We calculate the transverse momentum dependent
  gluon-to-gluon splitting function within $k_T$-factorization,
  generalizing the framework employed in the calculation of
  the quark splitting functions in~\cite{Hautmann:2012sh,
  Gituliar:2015agu, Hentschinski:2016wya} and demonstrate at the
  same time the consistency of the extended formalism with previous results. 
  While existing versions of $k_T$ factorized evolution 
  equations contain already a gluon-to-gluon splitting function {\it
    i.e.} the leading order Balitsky-Fadin-Kuraev-Lipatov (BFKL)
  kernel or the Ciafaloni-Catani-Fiorani-Marchesini (CCFM) kernel, the
  obtained splitting function has the important property that it
  reduces both to the leading order BFKL kernel in the high energy
  limit, to the Dokshitzer-Gribov-Lipatov-Altarelli-Parisi (DGLAP) 
  gluon-to-gluon splitting function in the collinear limit as well as
  to the CCFM kernel in the soft limit. At the same time we
  demonstrate that this splitting kernel can be obtained from a direct
  calculation of the QCD Feynman diagrams, based on a combined
  implementation of the Curci-Furmanski-Petronzio formalism for the
  calculation of the collinear splitting functions and the framework
  of high energy factorization.}

\end{abstract}

\newpage
{
  \hypersetup{linkcolor=black}
  \tableofcontents
}


\section{Introduction}
Parton distributions functions (PDFs) are crucial elements of collider
phenomenology. In presence of a hard scale $M$ with
$M \gg \Lambda_{\text{QCD}}$ and $\Lambda_{\text{QCD}}$ the QCD
characteristic scale of the order of a few hundred MeV, factorization
theorems allow to express cross-sections as convolutions of parton
densities and hard matrix elements, where the latter are calculated
within perturbative QCD~\cite{Collins:1984xc}.  This was first
achieved within the framework of collinear
factorization~\cite{Ellis:1978ty,Collins:1985ue,Bodwin:1984hc}, where the
incoming partons are taken to be collinear with the respective mother
hadron.  Calculating hard matrix elements to higher orders in the strong coupling constant,  one can systematically improve the precision of the
theoretical prediction, by incorporating more loops and more
emissions of real partons.  These extra emissions allow to
improve the kinematic approximation inherent to the leading order
description.
\\

As an alternative to improving the kinematic description through the
calculation of higher order corrections, one may attempt to treat
kinematics exactly from the very beginning, {\it i.e.} to account for
the bulk of kinematic effects already at leading order. An important
example of such kinematic effects is the transverse momentum $k_T$ of
the initial state partons, which is set to zero within collinear
factorization. Therefore, in collinear factorization, these effects
can only be recovered through the calculation of higher order
corrections, in particular through real parton emissions from initial
state particles, which generate finite transverse momenta for the
initial state partons of the observed hard event. Schemes which
provide an improved kinematic description already at the leading order
involve in general Transverse-Momentum-Dependent (TMD) or
`unintegrated' PDFs\footnote{For a review see
  \cite{Angeles-Martinez:2015sea}.}. \\

 TMD PDFs arise naturally in
regions of phase space characterized by a hierarchy of scales.  A
particularly interesting example is provided by the so called low $x$
region, where $x$ is the ratio of the hard scale $M^2$ of the process
and the center-of-mass energy squared $s$. The low $x$ region
corresponds therefore to the hierarchy $s \gg M^2 \gg
\Lambda_{\text{QCD}}^2$.  In such a kinematical setup, large
logarithms $\ln 1/x$ can compensate for the smallness of the
perturbative strong coupling $\alpha_s $ and it is necessary to resum
terms $\left(\alpha_s \ln 1/x \right)^n$ to all orders to maintain
the predictive power of the perturbative expansion.  Such a
resummation is achieved by the Balitsky-Fadin-Kuraev-Lipatov (BFKL)
\cite{Fadin:1975cb,Kuraev:1976ge,Kuraev:1977fs,Balitsky:1978ic}
evolution equation.  Its formulation is based on factorization of QCD
amplitudes in the high energy limit, $s \gg M^2$.  As a natural
by-product of such a factorization, one obtains QCD cross-sections as
convolutions in transverse momentum: in particular, cross-sections are
automatically factorized into $k_T$ dependent impact factors and the BFKL
Green's function.  Matching of high energy factorization to collinear
factorization which identifies properly normalized impact factors and
Green's function with unintegrated gluon density and $k_T$-dependent
perturbative coefficients is then achieved by so-called
$k_T$-factorization \cite{Catani:1990eg}; see
\cite{Deak:2009xt,vanHameren:2012if,vanHameren:2012uj,
  Chachamis:2015ona} for studies of various processes within this
scheme.
\\

While high energy factorization provides a well defined framework for
calculations of evolution kernels and coefficient functions also
beyond leading order, the applicability of the results is naturally
limited to the low $x$ limit of hard scattering events. If the ensuing
formalism is na\"ively extrapolated to intermediate or even large
values of $x$, the framework is naturally confronted with a series of
problems and short-comings. To name a few, contributions of quarks to
the evolution arise as a pure next-to-leading order (NLO) effect and
elementary vertices violate energy conservation {\it i.e.}
conservation of the hadron longitudinal momentum fraction.  While
these effects are subleading in the strict limit $x \to 0$, they
become sizeable if intermediate values of $x$ are reached.  The only way to
account for such effects within the high energy factorization
framework is through the determination of perturbative higher order
terms.  Apart from the direct calculation of perturbative higher order
corrections \cite{Fadin:1998py,Ciafaloni:1998gs}, this can be achieved
through including a resummation of terms, which restore subleading but
numerically relevant pieces of the
Dokshitzer-Gribov-Lipatov-Altarelli-Parisi (DGLAP)
\cite{Gribov:1972ri,Altarelli:1977zs,Dokshitzer:1977sg} splitting
functions
\cite{Salam:1998tj,Altarelli:2005ni,Ciafaloni:2003kd,SabioVera2005,Hentschinski:2012kr,Hentschinski:2013id};
for early attempts to unify the DGLAP and BFKL approaches see
\cite{Kwiecinski:1997ee,Ciafaloni:2003kd} as well as the more recent
attempt\cite{Bonvini:2017ogt}. Even though these resummations have
been successful in stabilizing low $x$ evolution into the region of
intermediate $x \sim 10^{-2}$, extrapolations to larger values of $x$ are still prohibited. 
Moreover, by merely resumming and calculating higher order corrections within the BFKL formalism, 
one essentially repeats the program initially outlined for collinear factorization:
higher order corrections are calculated to account for kinematic effects which are beyond the regarding factorization scheme. \\

To arrive at a framework which avoids the need to account for
kinematic effects through the calculation of higher order corrections,
it is therefore necessary to arrive at a scheme which accounts for
both DGLAP (conservation of the longitudinal momentum fraction) and
BFKL (conservation of the transverse momentum) kinematics. Note that
the mere definition of such a scheme is difficult: neither the hard
scale of the process (as in the case of DGLAP evolution) nor $x$ (BFKL
evolution) provides at first a suitable expansion parameter, if one
desires to keep exact kinematics in both variables. To overcome these
difficulties, we follow here a proposal initially outlined
in~\cite{Catani:1994sq}: There, following the all-order definition of
DGLAP splitting function within the Curci-Furmanski-Petronzio
(CFP)~\cite{Curci:1980uw} formalism, the low $x$ resummed DGLAP
splitting functions have been constructed. As an ingredient for such a
resummation program, the authors of~\cite{Catani:1994sq} were able to
define a TMD gluon-to-quark splitting function $\tilde{P}_{qg}$, both
exact in transverse momentum and longitudinal momentum fraction
(hereafter, we will use the symbol $\tilde{P}$ to indicate a
transverse momentum dependent splitting function).  While the TMD
gluon-to-quark splitting function is well defined within the low $x$
resummation program of~\cite{Catani:1994sq}, the same is not true for
the other splitting kernels.  In \cite{Hautmann:2012sh} it was then
demonstrated that the gluon-to-quark splitting function
$\tilde{P}_{qg}$ can be re-obtained through a simple extension of high
energy factorized amplitudes to exact kinematics. Following this
observation, two of us calculated in \cite{Gituliar:2015agu} the
remaining splitting functions which involve quarks, $\tilde{P}_{gq}$,
$\tilde{P}_{qg}$ and $\tilde{P}_{qq}$.  In this paper we continue this
program and compute the still missing TMD gluon-to-gluon splitting
function $\tilde{P}_{gg}$.  This requires a further modification of
the formalism used in~\cite{Catani:1994sq,Gituliar:2015agu}, in
particular when defining proper projection operators for high energy
factorization. As a result we obtain a TMD gluon-to-gluon splitting
function which agrees in the regarding limits not only with the DGLAP
and BFKL limits, but also reduces in the soft limit to the
Ciafaloni-Catani-Fiorani-Marchesini
(CCFM)~\cite{Ciafaloni:1987ur,Catani:1989yc,CCFMd} splitting kernel.
\\

The outline of this paper is as follows. In Sec.~\ref{real}, we recall
the definition of the TMD splitting functions and provide details
about the steps for the calculation of their real contributions.
Sec.~\ref{vertices} establishes an extension of the method previously
used to derive gauge invariant 3-point vertices, and uses it to
compute the corresponding 3-gluon vertex; the former used Lipatov's
effective action, the latter resorts to spinor helicity techniques.
Sec.~\ref{sec:proj} is dedicated to a discussion of projection
operators and their necessary modifications compared to
refs.~\cite{Curci:1980uw,Catani:1994sq}.  Sec.~\ref{sec:results}
contains the central results of this paper {\it i.e.} the complete set of
real contributions to the TMD splitting functions.
Sec.~\ref{sec:conclusions} is dedicated to a discussion of our
results.  Two appendices~\ref{AppSpinors} and~\ref{app:results}
contain supplementary details and a representation of splitting
kernels using an alternative set of variables.

\section{Definition of TMD splitting functions: real contributions}
\label{real}

\begin{figure}[ht]
\begin{center}
\subfloat[$\tilde{P}_{qg}$\label{fig:pqg}]{
\includegraphics[scale=1.0]{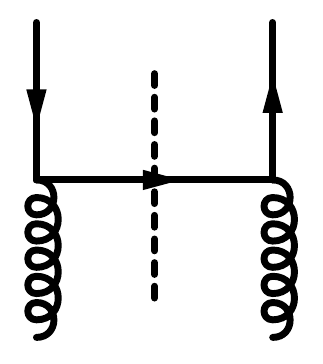}}
\subfloat[$\tilde{P}_{gq}$\label{fig:pgq}]{
\includegraphics[scale=1.0]{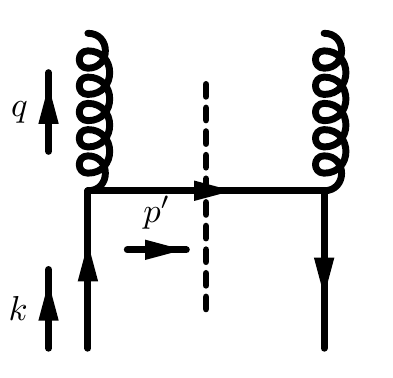}}
\subfloat[$\tilde{P}_{qq}$\label{fig:pqq}]{
\includegraphics[scale=1.0]{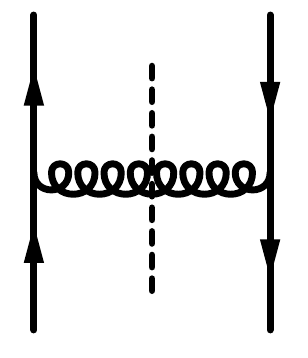}}
\subfloat[$\tilde{P}_{gg}$\label{fig:pgg}]{
\includegraphics[scale=1.0]{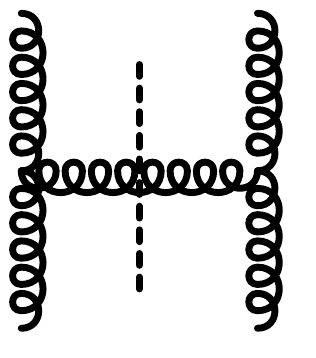}}
\caption{Squared matrix elements for the determination of the real contributions to the splitting functions \`{a}-la Curci-Furmanski-Petronzio.
Lower (incoming) lines carry always momentum $k$, upper (outgoing) lines carry momentum $q$.}
\label{real_graphs}
\end{center}
\end{figure}

The matrix elements involved in the calculation of the real
contributions to the leading order (LO) splitting functions are
presented in Fig.~\ref{real_graphs}.  The incoming momentum, called
$k$, features high energy kinematics, while the outgoing momentum,
$q$, is taken in its most general form. The 4-momenta will be
parametrised as follows
\begin{align}
 \label{eq:kinematics}
&
k^\mu = y p^\mu + k_\perp^\mu \, ,
\quad
q^\mu = x p^\mu + q_\perp^\mu + \frac{q^2+\qt^2}{2x\sp{p}{n}} n^\mu \, ,
\quad
\qtt  = \qt - z \kt \,.
 \end{align}
 Here $p$ and $n$ are two light-like momenta
 $\left( p^2 = n^2 = 0 \right)$ which refer to the two different
 light-cone directions for a fixed scattering axis; in the case of
 Deep-Inelastic-Scattering, one would for instance parametrize the
 virtual photon momentum as $q = n - x p$ with
 $x = Q^2/(2 p \cdot q)$, while $p$ would yield the proton momentum in
 the limit of zero proton mass.  We will also use $z=x/y$ to denote
 the longitudinal momentum fraction of the initial parton $k$ carried
 on by the parton $q$.  Within this setup, pure high energy kinematics
 corresponds to $z=0$, while collinear kinematics is obtained for
 $\kt=0$.  Following the procedure outlined
 in~\cite{Catani:1994sq,Hautmann:2012sh,Gituliar:2015agu}, we start
 from the definition of the 2-particle irreducible (2PI) TMD kernel
\begin{equation}
\label{eq:TMDkernelDefINI}
\hat K_{ij} \left(z, \frac{\kt^2}{\mu^2}, \epsilon, \alpha_s \right) =
z\,  \int \frac{d q^2 d^{2 + 2 \epsilon} {\qt}}{2 (2 \pi)^{4 + 2\epsilon}}\, \Theta(\mu_F^2 + q^2)  \Pbb_{j,\,\text{in}} \otimes \hat{K}_{ij}^{(0)}(q, k) \otimes \Pbb_{i,\,\text{out}}\, ,
\end{equation}
where $\hat{K}_{ij}^{(0)}$, $i,j=q,g$ denotes the squared matrix
element for the transition of a parton $j$ to a parton $i$, see
Fig.~\ref{real_graphs}, which includes the propagators of the outgoing
lines.  $\Pbb_{j,\,\text{in}}$ and $\Pbb_{j,\,\text{out}}$ are
projectors, to be discussed in detail in Sec. \ref{sec:proj}.  Gluons
are taken in the $n\cdot A=0$ light-cone gauge $(n^2=0)$.  The symbol
$\otimes$ represents contraction of spin indices (see also
Sec.~\ref{sec:proj}); $\mu_F$ denotes the factorization scale, and we
use dimensional regularisation in $d=4+2\epsilon$ dimensions with
$\mu^2$ the dimensional regularisation scale.  It is convenient to
introduce the following notation for the matrix element convoluted
with the projectors,
\begin{equation}
g^2\, 2 \pi \, \delta\left((k-q)^2\right)\, W_{ij} = \Pbb_{j,\,\text{in}} \otimes \hat{K}_{ij}^{(0)}(q, k) \otimes \Pbb_{i,\,\text{out}} \, ,
\label{eq:Wdef}
\end{equation} 
with
\begin{align}
\label{eq:delta_qtt}
\delta\left((k-q)^2\right) &= \frac{z}{1-z}\;
\delta\left(q^2 + \frac{\qtt^2 + z(1-z)\kt^2}{1-z}\right) \, .
\end{align}
With the $\overline{\text{MS}}$ strong coupling constant $\alpha_s  = \frac{g^2 \mu^{2\epsilon} e^{\epsilon \gamma_E}}{(4 \pi)^{1+ \epsilon}}$,
and  $\gamma_E$  the Euler-Mascheroni constant,
\begin{equation}
\frac{g^2}{2 (2 \pi)^{3 + 2\epsilon}} = 
\frac{\alpha_s}{4\pi}
\frac{e^{-\epsilon\gamma_E}}{\pi^{1+\epsilon}\mu^{2\epsilon}} \, ,
\end{equation}
and 
$ \qtt=\qt-z\kt$, 
 Eq.~(\ref{eq:TMDkernelDefINI})  turns into 
\bea
\label{eq:KijTMP2}
\hat K_{ij} \left(z, \frac{\kt^2}{\mu^2}, \epsilon, \alpha_s \right) 
&=&
z\, \frac{\alpha_s}{4\pi}\, \frac{e^{-\epsilon\gamma_E}}{\mu^{2\epsilon}} \int \frac{d^{2+2\epsilon}\qtt}{\pi^{1+\epsilon} \, \qtt^2} \,
\frac{z}{1-z}\, \qtt^2\, W_{ij}\Big|_{q^2=-\frac{\qtt^2 + z(1-z)\kt^2}{1-z}}
\nn \\
&\times&
\Theta\left(\mu_F^2 - \frac{\qtt^2 + z(1-z)\kt^2}{1-z}\right) \, .
\eea
This allows us to  identify the transverse momentum dependent splitting function $\tilde{P}_{ij}^{(0)}$ as
\beq
\tilde{P}_{ij}^{(0)} (z, \qtt, \kt ) =
\frac{z}{1-z}\, \qtt^2\, \frac{1}{2} W_{ij}\Big|_{q^2=-\frac{\qtt^2 + z(1-z)\kt^2}{1-z}} \; ,
\eeq 
and we obtain
\beq
\label{eq:KijangDep}
\hat K_{ij} \left(z, \frac{\kt^2}{\mu^2}, \epsilon, \alpha_s \right)  =
\frac{\alpha_s}{2\pi} \, z\,  \frac{e^{-\epsilon\gamma_E}}{\mu^{2\epsilon}}
\int \frac{d^{2+2\epsilon}\qtt}{ \pi^{1+\epsilon} \,\qtt^2} \, \tilde{P}_{ij}^{(0)} \,
\Theta\left(\mu_F^2 - \frac{\qtt^2 + z(1-z)\kt^2}{1-z}\right) \, .
\eeq
With the  angular averaged TMD splitting function  defined as
\begin{align}
\label{eq:Pij}
\bar{P}_{ij}^{(0)} = 
\frac{1}{\pi} \int_0^{\pi} d\phi \,\sin^{2\epsilon}\phi \;
\tilde{P}_{ij}^{(0)}
\end{align}
we finally  arrive at  
\begin{align}
\hat K_{ij} \left(z, \frac{\kt^2}{\mu^2}, \epsilon, \alpha_s \right) &=
\frac{\alpha_s}{2\pi} z
\frac{e^{-\epsilon\gamma_E}}{\Gamma(1+\epsilon)}
\frac{1}{2}
\int_0^{(1-z)(\mu_F^2-z\kt^2)} \frac{d\qtt^2}{\qtt^2} \,
\left(\frac{\qtt^2}{\mu^{2}}\right)^\epsilon
\bar{P}_{ij}^{(0)}\left(z, \frac{\kt^2}{\qtt^2} \right) \, .
\end{align}

\section{Production vertices from spinor helicity amplitudes}
\label{vertices}

The calculation of the real contributions to the $\tilde{P}_{qg}$,
$\tilde{P}_{gq}$ and $\tilde{P}_{qq}$ in ~\cite{Gituliar:2015agu} was
based on the effective 3-point vertices,
\bea
\Gamma^\mu_{q^*g^*q}(q,k,p') 
&=&
i\,g\,t^a\, \left( \gamma^\mu - \frac{n^\mu}{k\cdot n}\, \slashq  \right) ,
\label{Lipatov_vertices_1} \\
\Gamma^\mu_{g^*q^*q}(q,k,p') 
&=&
i\,g\,t^a\, \left(  \gamma^\mu - \frac{p^\mu}{p\cdot q}\, \slashk \right),
\label{Lipatov_vertices_2} \\
\Gamma^\mu_{q^*q^*g}(q,k,p') 
&=&
i\,g\,t^a\, \left( \gamma^\mu - \frac{p^\mu}{p\cdot p'}\, \slashk + \frac{n^\mu}{n\cdot p'}\, \slashq  \right).
\label{Lipatov_vertices_3}
\eea
These vertices have been obtained from Lipatov's effective action
formalism~\cite{Lipatov:1995pn,Antonov:2004hh} and afterwards slightly
generalized to the TMD kinematics of Eq.~\eqref{eq:kinematics}.  While
a corresponding vertex $\Gamma_{g^*g^*g}$ can be easily obtain from
Lipatov's effective action, the generalization to TMD kinematics turns
out to be far less trivial. Within the effective action formalism, an
off-shell gluon corresponds to a reggeized gluon which is
automatically associated with a specific polarization, proportional to
the light-cone momenta $p$ and $n$. While this is sufficient for the
incoming off-shell gluon with momentum $k$, the CFP formalism requires
open indices for the out-going gluon with momentum $q$. One is
therefore driven to consider instead, the so-called
gluon-gluon-reggeized gluon (GGR) vertex. This vertex is well known,
see \cite{Antonov:2004hh,Chachamis:2012cc} for a construction in
covariant gauges.  Indeed we will find that use of the corresponding
GGR vertex in $A \cdot n= 0$ light-cone gauge is sufficient to
calculate the TMD splitting kernel.  Nevertheless the direct use of
this vertex is not completely satisfactory: Within this vertex, the
gluon with momentum $q$ is treated as an ordinary QCD gluon;
off-shellness of this gluon leads then naturally to a violation of
current conservation and therefore gauge invariance. Below we verify
that current conservation and therefore gauge invariance is restored
by adding a term proportional to $n^\mu$, which is set to zero within
the employed light-cone gauge. While such a restoration of current
conservation might have been expected from the very beginning, we
demonstrate below that such a term indeed arises out of a proper
Feynman diagram analysis, deviating slightly from the strategy
employed in~\cite{Gituliar:2015agu}. In particular we demonstrate that
the necessary production vertices can be as well obtained from a
direct study of QCD scattering amplitudes
in the high energy limit.\\

To this end we will first recover the vertices
(\ref{Lipatov_vertices_1})-(\ref{Lipatov_vertices_3}) by stripping off
the helicity dependence from scattering amplitudes computed by
applying spinor helicity methods to high energy factorization
~\cite{vanHameren:2012uj,vanHameren:2012if,vanHameren:2013csa,
  vanHameren:2014iua,vanHameren:2015bba,vanHameren:2016bfc}.  In a
next step we will use then this formalism to infer the structure of
the 3-point gluon vertex to be used to compute $\tilde{P}_{gg}(z)$.
While our result is obtained within the spinor helicity formalism, we
would like to stress that this formalism provides merely a convenient
framework for fast calculation; the obtained result is on the other hand
completely general and could have been equally obtained through a
study of QCD amplitudes using conventional Feynman rules. For an
exhaustive reviews of the spinor helicity formalism, we refer the
reader to~\cite{Dixon:1996wi,Mangano:1990by,Elvang:2013cua}.\\

While we finally aim at the TMD kinematics Eq.~\eqref{eq:kinematics},
the analysis in this section is at first limited to kinematics
required by high energy factorization (see below for the precise
parametrization used). The more general TMD kinematics
Eq.~\eqref{eq:kinematics} will then be addressed in
Sec.~\ref{sec:gauge-invar-effect}.  The reader who is only interested
in the final form of the vertices can skip the rest of this section
and just take for granted the formulae
(\ref{Lipatov_vertices_1})-(\ref{Lipatov_vertices_3}) for the 3-point
vertices with a fermion pair, as well as
the result (\ref{eq:ggg_vertex}) for the 3-gluon vertex. \\

To keep our derivation self contained, we recall the basic idea of the
spinor helicity method to derive gauge invariant scattering amplitudes
with off-shell particles~\cite{vanHameren:2012if,vanHameren:2013csa}:
each off-shell particle is introduced into the diagram through an
auxiliary on-shell pair; the squared sum of the on-shell momenta of
the auxiliary pair accounts for the non vanishing squared momentum
particle that they introduce.  In the case of a gluon, a
quark-antiquark pair is used; an off-shell quark is introduced via a
vertex featuring an auxiliary photon-quark pair.\footnote{As stressed
  in \cite{vanHameren:2013csa}, these auxiliary 'photons' interact
  only with the corresponding auxiliary fermion and not with any other
  fermion with electromagnetic charge; they can actually be thought of
  as bosons mediating an extra fictitious abelian interaction. Since
  the purpose of their introduction is to build an amplitude which is
  off-shell and gauge-invariant, and whose dependence from the
  auxiliary particles is eventually dropped, this is legitimate.}  The
price to pay is the proliferation of Feynman diagrams, which are more
than in the on-shell case; the additional diagrams feature propagators
of the auxiliary particles.  As demonstrated
in~\cite{vanHameren:2012if}, in the given kinematics the colour
degrees of freedom of the pairs of auxiliary quarks are exactly
equivalent to the colour of the corresponding off-shell gluon.  On the
other hand, since the auxiliary particles are on-shell, the gauge
invariance of the resulting scattering amplitude is immediately
manifest.

\subsection{Production by two off-shell quarks}

\begin{figure}[ht]
\begin{center}
\includegraphics[scale=0.80]{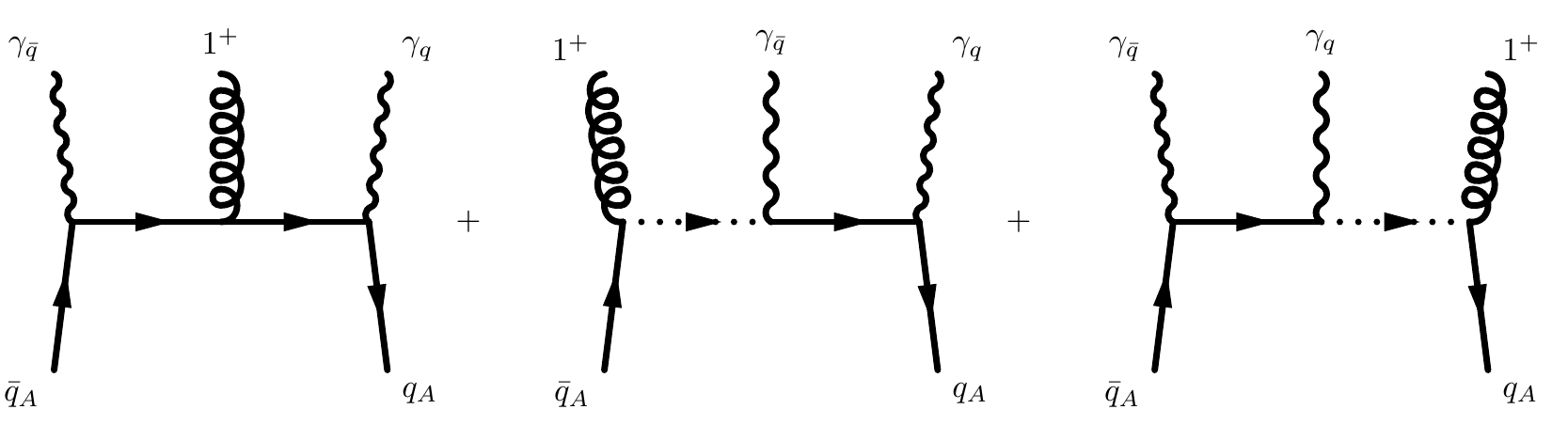}
\caption{Feynman diagrams contributing to $\Amp(1^+,\qb^{*+},q^{*-})$.}
\label{off_qqb}
\end{center}
\end{figure}

We start with the simplest case, the production of an on-shell gluon
by two off-shell fermions.  By construction, amplitudes in the spinor
helicity formalism are computed for specific values of the massless
particle helicities~\cite{Mangano:1990by}.  This implies that the
un-contracted off-shell vertex can be obtained from the amplitude by
going backwards from the final result and stripping off the helicity
dependence.  The Feynman diagrams using the rules
of~\cite{vanHameren:2013csa} are given in Fig~\ref{off_qqb}.
All the momenta are taken as incoming with
\beq
k_q = x_q\,n + k_{q\perp} \, , \quad k_{\qb} = x_{\qb}\, p + k_{\qb\perp} \, , \quad k_1 = p_1 \, , 
\quad
k_q + k_{\qb} + p_1 = 0 \,,
\label{mom_par_offqoffqg}
\eeq
while  $p_q = n$ and $p_{\qb} = p$. Note that $k_q = -q|_{n\cdot q=0}$ and $k_{\qb} = k$.  In the last two
diagrams of Fig. \ref{off_qqb}, the auxiliary fermions propagate
eikonally through the diagrams; this is indicated with a dashed line.
The resulting amplitude is then given by
\bea
\Amp(1^+,\qb^{*+},q^{*-}) 
&=& 
\AL{q}   
\frac{\slasheps_{q\, +}}{\sqrt{2}} \frac{-\slashk_q}{k_q^2}  \frac{\slasheps_{1\,+}}{\sqrt{2}} \frac{\slashk_{\qb}}{k_{\qb}^2}  \frac{\slasheps_{\qb\,-}}{\sqrt{2}}  +
\frac{\slasheps_{q\,+}}{\sqrt{2}} \frac{\slashk_q}{k_q^2}  \frac{\slasheps_{\qb\,-}}{\sqrt{2}} \frac{\slashp}{2 p\cdot k_q}  \frac{\slasheps_{1\,+}}{\sqrt{2}} +
\nn \\
&& \hspace{6mm}
\frac{\slasheps_{1\,+}}{\sqrt{2}} \frac{\slashn}{2 n\cdot k_{\qb}}  \frac{\slasheps_{q\,+}}{\sqrt{2}} \frac{\slashk_{\qb}}{k_{\qb}^2}  \frac{\slasheps_{\qb\,-}}{\sqrt{2}} 
\SR{\qb} \, .
\eea
In the first step we  remove the polarization vector of the gluon 
$\epsilon_{1+}^{\mu}/{\sqrt{2}}$,\footnote{The proliferation of $\sqrt{2}$ factors is typical for the colour-ordered Feynman rules~\cite{vanHameren:2014iua,Dixon:1996wi}.}
in order to get rid of the dependency on the helicity of the on-shell gluon, and arrive at 
\beq
\Amp(1^+,\qb^{*+},q^{*-}) \rightarrow 
\frac{1}{2}\, \AL{q} -
\frac{ \slasheps_{q\, +} \, \slashk_q \, \gamma^\mu\,  \slashk_{\qb}\,  \slasheps_{\qb\,-}}{k_q^2\, k_{\qb}^2}    +
\frac{ \slasheps_{q\,+} \, \slashk_q \, \slasheps_{\qb\,-} \, \slashp \, \gamma^\mu}{k_q^2\, 2 p\cdot k_q} +
\frac{\gamma^\mu \, \slashn \, \slasheps_{q\,+} \, \slashk_{\qb} \, \slasheps_{\qb\,-} }{ 2 n\cdot k_{\qb} \, k_{\qb}^2}
\SR{\qb} \, .
\label{gqboffqoff}
\eeq
As the auxiliary photons are not observable, one should not keep track of their helicities, summing over them at the amplitude level. 
Nevertheless it turns out that  there is  only one choice  which does
not return a vanishing result. In particular,  we can choose the momenta of the fermion
pair, $n$ and $p$, to be the auxiliary momenta necessary to build 
the polarization vector of each other's auxiliary photon, {\it i.e.}
\beq
\epsilon^\mu_{q\,+} = \frac{\AL{\qb} \gamma^\mu \SR{q}}{ \sqrt{2} \AA{\qb q}} \, , \quad 
\epsilon^\mu_{\qb\,-} = \frac{\AL{\qb} \gamma^\mu \SR{q}}{ \sqrt{2} \SSS{\qb q}} \, .
\label{offqoffqg_pol}
\eeq
In this way, when we explicitly open the slashed polarization vectors by using the spinor identity
\beq
\AL{a} \gamma^\mu c_\mu \SR{b} = 2 \, \left(  \AR{a}\SL{b} + \SR{b}\AL{a} \right)\, ,
\eeq
the first and third term  of   Eq.~\eqref{gqboffqoff} vanish because $\AL{p}\AR{p} = \SL{p}\SR{p} = 0$ for any massless momentum $p$ 
and  we obtain
\beq
- 
\frac{\AL{\qb} \slashk_{\qb} \, \gamma^\mu \, \slashk_{q}  \SR{q}}{k_q^2\,k_{\qb}^2} + 
\frac{\AL{\qb}\slashk_q \SR{q} \, p^\mu_{\qb}}{k_q^2 \, p_{\qb}\cdot k_{q}} + 
\frac{\AL{\qb}\slashk_{\qb} \SR{q} \, p^\mu_{q}}{k_{\qb}^2 \, p_{q}\cdot k_{\qb}}  \, .
\eeq
This can be conveniently reshuffled as follows,
\beq
\AL{\qb} \frac{\slashk_{\qb}}{k_{\qb}^2} \, \left\{
\gamma^\mu + \frac{p^\mu}{p\cdot k_q} \slashk_{\qb} + \frac{n^\mu}{n\cdot k_{\qb}} \slashk_q
\right\} \, \frac{\slashk_q }{k_q^2} \SR{q} \, ,
\eeq
and one can easily check that the term in curly brackets coincides
with   the vertex of Eq.~(\ref{Lipatov_vertices_3}) for the high
energy kinematics Eq.~\eqref{mom_par_offqoffqg}.

\subsection{Production by  an off-shell quark and  an off-shell  gluon}

\begin{figure}[ht]
\begin{center}
\includegraphics[scale=0.5]{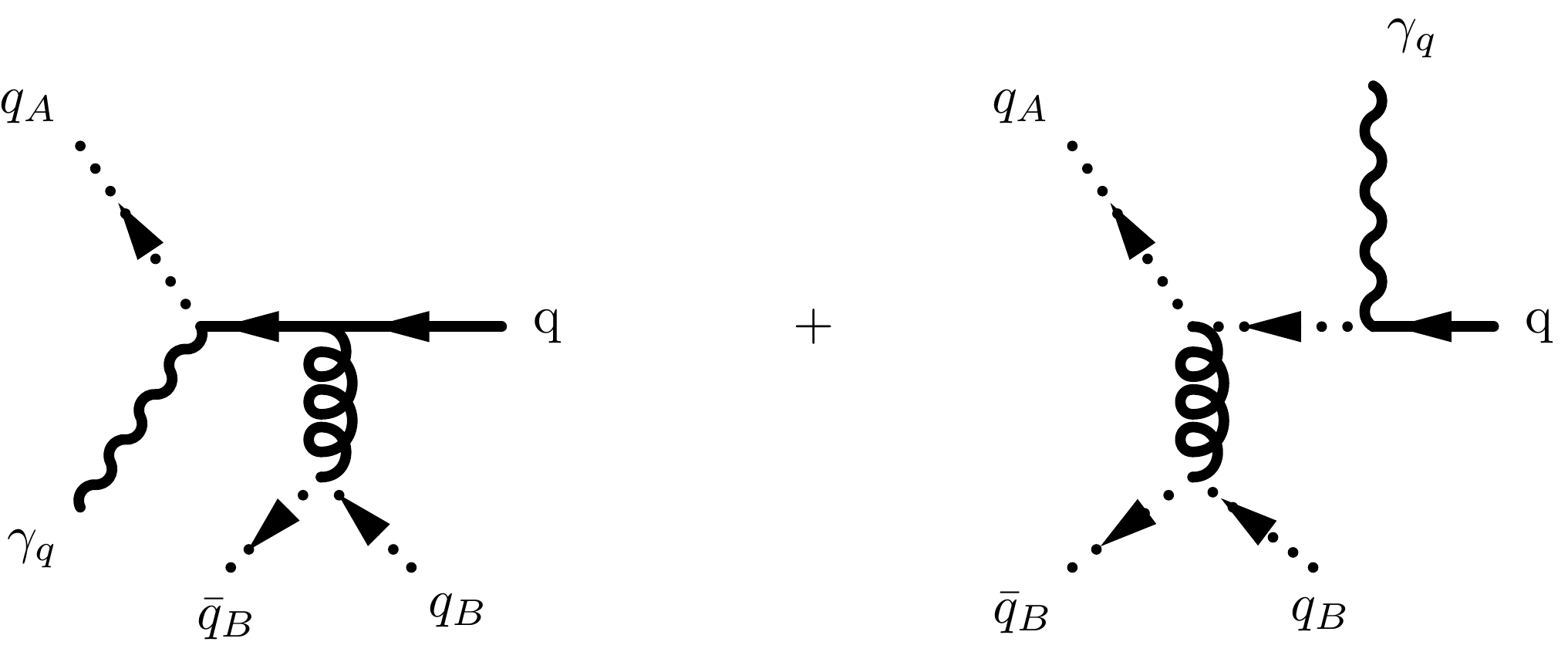}
\caption{Feynman diagrams contributing to $\Amp(1^*,\qb^{*+},q^-)$.}
\label{off_gqb}
\end{center}
\end{figure}

In the cases with one  off-shell gluon, the contributing Feynman diagrams are shown in Fig.~\ref{off_gqb}. 
The gluon propagator with momentum $k$ will be now taken in  light-cone gauge as
\beq
\label{eq:gammamu_g-q-q}
\frac{d_{\mu\nu}(k)}{k^2} \, , \quad \text{with} \quad d_{\mu\nu}(k) = -g_{\mu\nu} + \frac{k_\mu n_\nu + k_\nu n_\mu}{k\cdot n}  \, .
\eeq
Again, all the momenta are taken as incoming and their parametrisation is
\beq
k_1 = x_1\,n + k_{1\perp} \, , \quad k_{\qb} = x_{\qb}\, p + k_{\qb\perp} \, , \quad k_q = p_q \, ,
\quad
\quad
k_1 + p_q + k_{\qb} = 0 \, ,
\label{mom_par_offqboffgq}
\eeq
while $p_{\bar{q}} = p$  and $p_1 = n$; we also have  $k_1 =- q|_{n \cdot q = 0}$ while $ k_{\qb} = k$.
We find for the amplitude
\beq
\Amp(1^*,\qb^{*-},q^{+}) =
\frac{\AL{1}\gamma^\mu\SR{1}}{\sqrt{2}}\, \frac{d_{\mu\nu}(k_1)}{k_1^2} \, 
\AL{\qb} 
\frac{\slasheps_{\qb +}}{\sqrt{2}} \frac{\slashk_{\qb}}{\qb^2} \frac{\gamma^\nu}{\sqrt{2}} -
\frac{\gamma^\nu}{\sqrt{2}} \frac{\slashp}{2p_q\cdot p}\, \frac{\slasheps_{\qb +}}{\sqrt{2}}
\SR{q} \, .
\eeq
After inserting the explicit expressions for the polarization vectors of the auxiliary photon (see Appendix \ref{AppSpinors}), we obtain
\beq
\Amp(1^*,\qb^{*-},q^{+}) =
p_1^\mu \, \frac{d_{\mu\nu}(k_1)}{k_1^2}\, \SL{q} \, \left\{ \gamma^\nu - \frac{p^\nu}{p \cdot p_q}\, \slashk_{\qb} \right\} \, \frac{\slashk_{\qb}}{k_{\qb}^2} \SR{\qb} \, ,
\eeq
where the curly brackets contain the exact analogous of
Eq.~(\ref{Lipatov_vertices_2}).  The derivation of
Eq.~(\ref{Lipatov_vertices_1}) follows then immediately. 
Note that the polarization tensor of the gluon propagator  provides an
overall factor. 
If this result had been obtained in a covariant gauge, this numerator would be
\beq
d^{\text{cov.}}_{\mu\nu}(k_1) = - g_{\mu\nu} + \xi \frac{k_{1\,\mu} k_{1\,\nu}}{k_1^2} \, .
\eeq
The contraction of the vertex in curly brackets with $k_{1\nu}$
vanishes, as can be easily checked by momentum conservation and the
Dirac (Weyl) equation expressed in spinor helicity language $\SL{q} \slashp_q = 0$, 
so that only $-g_{\mu\nu}$ survives for a covariant gauge. 
This makes it irrelevant (up to a sign) to include or not the numerator in the factor outside the curly brackets. 
In the light-cone gauge, this is no longer the case, since the polarization tensor Eq.~\eqref{eq:gammamu_g-q-q}
is not invertible. 
Thus, the contraction of $n_\nu$ with the bracketed vertex does not
vanish and the equivalence with the covariant gauge is lost, at the
pure amplitude level. This does not have physical consequences, since
the amplitude is not an observable.  It might
however lead to  an ambiguity for the computation of the splitting function,
because the two expressions for the off-shell vertex,
\beq
\left\{ \gamma^\nu - \frac{p_{\qb}^\nu}{p_{\qb} \cdot p_q}\, \slashk_{\qb} \right\} 
\quad \text{vs.} \quad
\frac{d_{\mu\nu}(k_1)}{k_1^2}\, \, \left\{ \gamma^\nu - \frac{p_{\qb}^\nu}{p_{\qb} \cdot p_q}\, \slashk_{\qb} \right\} \, ,
\eeq
could lead to different splitting functions, when the CFP procedure is applied~\cite{Curci:1980uw,Gituliar:2015agu}.
Nevertheless such an ambiguity does not arise. %
First, because including the numerator of the gluon propagator is
necessary in order to be consistent (as we will demonstrate in the next section).
Second, because, when working completely consistently in light-cone gauge, 
a modified spin projector for the incoming gluons is necessary in order to apply 
the CFP procedure to off-shell matrix elements, as we will discuss extensively in Sec.~\ref{sec:proj}.
This completely eliminates any ambiguity and yields a result identical to the one obtained previously
for $\tilde{P}_{qg}$ by using the vertex without the numerator of the gluon propagator~\cite{Gituliar:2015agu}.

\subsection{Production by two off-shell gluons}

%
\begin{figure}[ht]
\begin{center}
\includegraphics[scale=0.8]{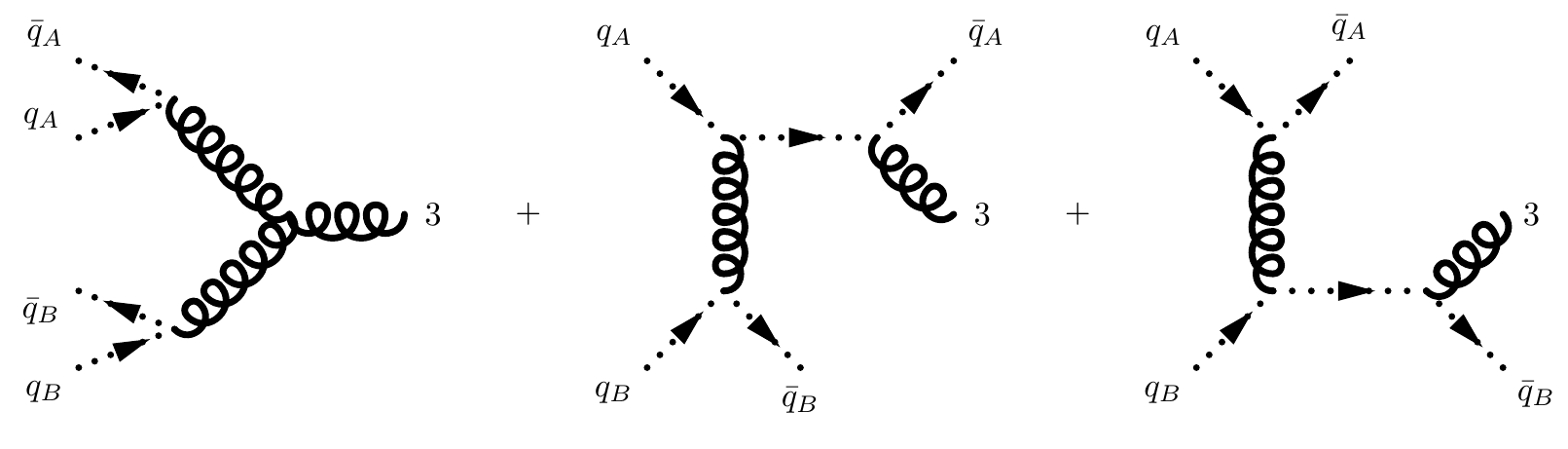}
\caption{Feynman diagrams contributing to $\Amp(1^*,2^*,3^{\pm})$. Note that we have only three diagrams instead of the usual five diagrams found in the derivation of the Lipatov vertex. This is merely due to color ordering imposed on the helicity amplitudes and does not affect the final result.}
\label{off_gg}
\end{center}
\end{figure}
The three Feynman diagrams for an amplitude with 2 off-shell and one
on-shell gluons are depicted in Fig.~\ref{off_gg}: the first piece is
the ordinary QCD 3-gluon vertex augmented by the propagators of the
off shell gluons with the corresponding couplings to the auxiliary
fermion pairs; 
 then there are the two extra contributions in which the
auxiliary quark lines eikonalize. The expression for the off-shell 3-gluon amplitude in terms of the three depicted contributions 
is given in~\cite{vanHameren:2014iua,Kotko:2014aba}.  Here we provide
only the result 
after factoring out the denominators of the off-shell gluon propagators
with the momenta
\beq
k_1 = x_1\,p + k_{1\perp} \, , \quad k_{2} = x_2\, n + k_{2\perp} \,  ,
\quad
k_1 + k_2 + p_3 = 0 \, ,
\label{eq:mom_par_offgoffgg}
\eeq
 and  $\mathcal{V}^{\mu_1\mu_2\mu_3}(k_1,k_2,p_3)$  the ordinary QCD
 three-gluon vertex, one has
\bea
\Amp(g_1^*,g_2^*,g_3) 
&=&
\sqrt{2}\, \frac{p_{ \mu_1} \,n_{\mu_2}\, \epsilon_{\mu_3}(p_3)}{k_1^2 \, k_2^2}\, \bigg\{ 
\mathcal{V}^{\lambda \kappa \mu_3}(k_1,k_2,p_3) \, {d^{\mu_1}}_{\lambda} (k_1)\, {d^{\mu_2}}_{\kappa}(k_2) 
\nn \\
&&\qquad \qquad \qquad 
+\, d^{\mu_1\mu_2}(k_2)\, \frac{k_1^2 p^{\mu_3}}{p\cdot p_3} -
d^{\mu_1\mu_2}(k_1)\, \frac{k_2^2 n^{\mu_3}}{n\cdot p_3}
\bigg\} \, .
\label{offgoffg_g}
\eea In~\cite{vanHameren:2014iua} the computations were performed in
the Feynman gauge, with $d^{\text{cov.}}_{\mu\nu}(k) = - g_{\mu\nu}$.
Here we use instead the light-cone gauge polarization tensor
Eq.~\eqref{eq:gammamu_g-q-q}.  Due to gauge invariance, the only
difference in the construction of the vertex is a change in the gluon
polarization tensor. Note that since the light-cone gauge polarization
tensor $d_{\mu\nu}(k)$, Eq.~\eqref{eq:gammamu_g-q-q}, is not
invertible, it is not possible to extract the numerators of both gluon
propagators simultaneously for all the three terms in the curly
brackets in Eq.~(\ref{offgoffg_g}).  We therefore retain these
polarization tensors in the expression for the off-shell vertex.\\

The above vertex, extracted  for high energy kinematics
Eq.~\eqref{eq:mom_par_offgoffgg} is then
used  to construct the corresponding vertex for TMD kinematics Eq.~\eqref{eq:kinematics}.  
The final form of the gluon production vertex in high energy kinematics,
using the naming convention for momenta as in
Eqs.~(\ref{Lipatov_vertices_1}-\ref{Lipatov_vertices_3}) and~\eqref{eq:kinematics},
is given by 
\bea
\label{eq:ggg_vertex} 
\Gamma^{\mu_1\mu_2\mu_3}_{g^*g^*g}(q,k,p') &=& \mathcal{V}^{\lambda
  \kappa \mu_3}(-q,k,-p') \, {d^{\mu_1}}_{\lambda} (q)\,
{d^{\mu_2}}_{\kappa}(k)
\nn \\
&& \qquad \qquad \qquad +\, d^{\mu_1\mu_2}(k)\, \frac{q^2 n^{\mu_3}}{ n\cdot p'} -
d^{\mu_1\mu_2}(q)\, \frac{k^2 p^{\mu_3}}{ p\cdot p'} \, ,
\notag \\
\text{with} && n\cdot k=n \cdot p' \qquad p\cdot q = -p\cdot p' \, .
\eea
As indicated in the last line of the above expression, this result is
so far only valid for the special case $n\cdot k=n \cdot p'$ since
$q \cdot n = 0$ in high energy kinematics. TMD kinematics
Eq.~\eqref{eq:kinematics} requires on the other hand
$ q \cdot n \neq 0$ which breaks the symmetry of the second term of
Eq.~\eqref{eq:ggg_vertex} under exchange of momenta
$k \leftrightarrow p'$.  We will find that the form already given in
Eq.~\eqref{eq:ggg_vertex} provides finally the correct generalization
to TMD kinematics (since it guarantees current conservation for the
produced gluon).  However, in order to demonstrate this we need to
make use of the structure of the modified projectors, which will be
discussed in the following section.

\section{Modifying the collinear projectors}
\label{sec:proj}

As a next step, we introduce the CFP approach to collinear factorization, explain the role of the projectors and 
motivate the need for the modifications which allow to consistently accommodate them in our framework.
To make the present discussion as self-contained as possible, we first remind the basics which motivated their introduction in the case 
of  collinear factorization~\cite{Ellis:1978sf,Curci:1980uw}, following closely Sec.~2.2 of~\cite{Catani:1994sq}.

\subsection{Collinear case}
%
\begin{figure}[t]
\begin{center}
\includegraphics[scale=0.5]{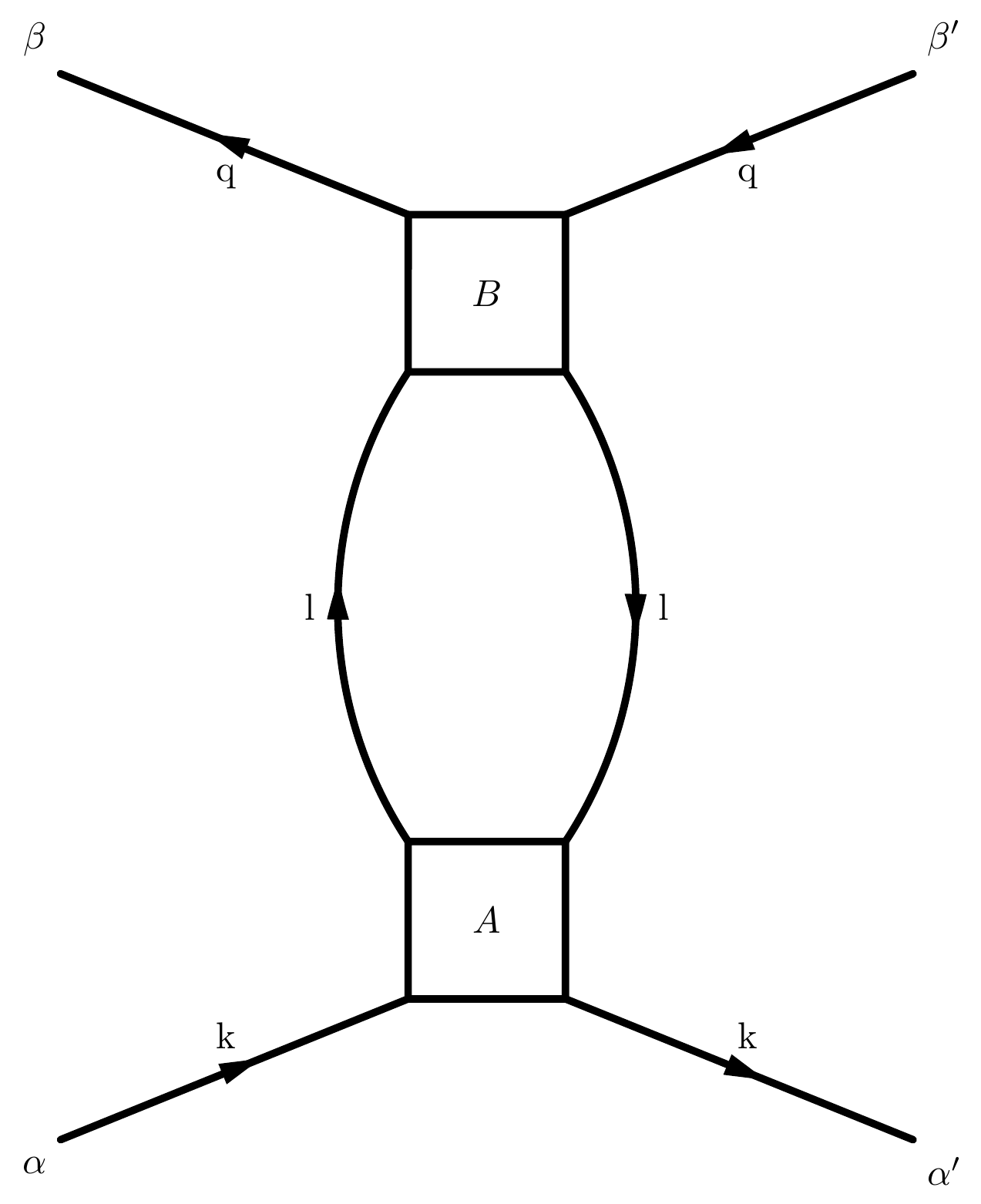}
\caption{Convolution of two generic 2PI amplitudes $A$ and $B$. Notice that the indices $\alpha$, $\beta$ and $\gamma$
maybe spinor as well as Lorentz indices, depending on the particles going in and coming out of the kernels.
The whole kernel, which  we can denote by $C$, is given by the convolution 
$C^{\alpha\alpha'}_{\beta\beta'}(k,q) = \int \frac{d^n l}{(2\pi)^n}\, A^{\alpha\alpha'}_{\gamma\gamma'}(k,l)\, B^{\gamma\gamma'}_{\beta\beta'}(l,q)$.}
\label{2PI_fig}
\end{center}
\end{figure}
The underlying concept  is based on the observation  that the hard matrix element contains collinear divergences which can be factorised,
to all orders in $\alpha_s$, and reabsorbed into a process-independent  transition-functions $\Gamma$.
Calling $\sigma^{(0)}$ the leading-twist contribution to the inclusive cross section, this can be schematically written as 
\beq
\sigma^{(0)} = C\, \Gamma \, ,
\label{leadtwist}
\eeq
where $C$ is the renormalized and finite (for $\epsilon \rightarrow 0$) hard matrix element,
$\Gamma$ is the transition function which absorbs only the collinear poles in $1/\epsilon$ and a momentum integration is implied.
The transition function can then be used to introduce the parton density functions,
\beq
\tilde{f}(x,\mu_F) \equiv \Gamma \, \tilde{f}^{(0)}(x,\mu_F) \, ,
\eeq
leading to the usual factorisation formula. The possibility to perform
such a factorisation was proven in~\cite{Ellis:1978sf}
and~\cite{Curci:1980uw},  
where use of the light-cone gauge turned out to be crucial.\footnote{See~\cite{Collins:2011zzd} for a formulation of the corresponding program which does not require light-cone gauges.}
The first result of~\cite{Ellis:1978sf} is that the cross section can be expanded as a series of 2PI 
(2-Particle-Irreducible) kernels $K^{(0)}$, describing the parton evolution
before the hard scattering, times a hard scattering coefficient function $C^{(0)}$,
\bea
\sigma^{(0)} 
&=& 
C^{(0)} \, \bigg( 1 + K^{(0)} + K^{(0)}\, K^{(0)} + \dots   \bigg) \equiv C^{(0)} \, \mathcal{G}^{(0)} \, ,
\nn \\
\mathcal{G}^{(0)} 
&\equiv& 
\bigg( 1 + K^{(0)} + K^{(0)}\, K^{(0)} + \dots   \bigg) = \frac{1}{1- K^{(0)}}\, .
\label{2PIexpansion}
\eea
By definition, each 2PI kernel $K^{(0)}$ (see Fig. \ref{2PI_fig} 
for an illustration of two linked 2PI kernels) has propagators only in the upper legs, 
which are connected by a momentum integral, $\int d^d l$, with the lower legs of the following 2PI kernel
all the way up to the coefficient function.
The central result of~\cite{Ellis:1978sf} is that the kernels themselves are free from infrared singularities in
the light-cone gauge, such that all the collinear poles in $1\over \epsilon$ 
come from the integration over the intermediate momenta connecting the kernels.
Then, one introduces a projector $\Pbb_C$ which isolates the infrared poles of $K^{(0)}$~\cite{Curci:1980uw},
\beq
K^{(0)} = \left(  1 - \Pbb_C \right)\, K^{(0)} + \Pbb_C\, K^{(0)} \, ,
\label{splitpoles}
\eeq
in such a way that the second term contains all the singularities, whereas the first is free of poles in $\epsilon$.
Factorization emerges when one notices that (\ref{splitpoles}) can be used recursively to reorganise 
the expansion of the process-independent Green function (\ref{2PIexpansion}) as
\begin{align}
\label{mathcalG} 
\mathcal{G}^{(0)} &= \mathcal{G} \, \Gamma \, , \qquad
\mathcal{G}   = \frac{1}{1-(1-\Pbb_C)\,K^{(0)} }  \, , 
\\
\label{gamm}
\Gamma &= \frac{1}{1- \Pbb_C \,K } = 1 +  \Pbb_C \,K + \left(\Pbb_C \, K\right) \, \left(\Pbb_C \,K \right) + \dots
\, , \quad K \equiv K^{(0)} \, \mathcal{G} \, ,
\end{align}
Since (\ref{mathcalG}) is clearly free of singularities, it is apparent that we have sketched how to arrive  at Eq.~(\ref{leadtwist}), setting 
\beq
C \equiv C^{(0)} \, \mathcal{G} \, .
\eeq
To further understand the projectors, we split them into one projector acting on the spin indices 
and one acting on the momentum space integral between two consecutive kernels,
\beq
\Pbb_C \equiv \Pbb^{\epsilon} \otimes \Pbb^{s} \, .
\eeq
Here, the momentum space projector $\Pbb^{\epsilon}$ extracts the singular part of the integral
over the intermediate momentum, transforming it into a convolution integral,
whereas $\Pbb^{s}$ decouples lower and upper kernel in the spin indices.
It is apparent that $\Pbb^{\epsilon}$ is independent of the particle species propagating through the 2PI ladder, 
whereas $\Pbb^{s}$ is closely related to it.  
In~\cite{Curci:1980uw} an explicit argument is presented which identifies the proper projector
for the non-singlet sector, whereas~\cite{Furmanski:1980cm} presents
the projector if two 2PI kernels are connected by a gluon.  
Referring to Fig.~\ref{2PI_fig} for our nomenclature of the kernels and the momenta, we have
\beq
A(q, l)_{\dots\alpha\alpha'}\, \Pbb^{s\, \alpha\alpha'}_{\,\,\, \beta\beta'}\, B^{\dots\beta\beta'}(l,k) \equiv
A(q, l)_{\dots\alpha\alpha'}\, \frac{(\slashl)^{\alpha\alpha'}}{2}\, \frac{(\slashn)_{\beta\beta'}}{2n\cdot l}\, B^{\dots\beta\beta'}(l,k)\, , 
\eeq
when the intermediate particle is a quark, with indices belonging to $\gamma$-matrices, and
\beq 
A(q,l)_{\dots\mu'\nu'}\, \Pbb^{s\, \mu'\nu'}_{\,\,\, \mu\nu}\,
B^{\dots\mu\nu}(l,k) \equiv A(q, l)_{\dots\mu'\nu'}\,
\frac{d^{\mu'\nu'}(l)}{d-2}\, (- g_{\mu\nu})\, B^{\dots\mu\nu}(l,k) \, ,
\eeq
if the intermediate particle is a gluon with Lorentz tensor
indices; $d$ specifies the number of space-time dimensions. 
We can further split the spin projectors into an "in" and "out" component
\begin{align}
\label{eq:3}
\Pbb^{s} \equiv \Pbb^{s}_{\text{in}} \otimes \Pbb^{s}_{\text{out}} \, .  
\end{align}
The names "in" and "out" can be understood from Fig.~\ref{2PI_fig} and interpreting the diagram in terms of a parton evolution
unfolding upwards. In other words, the amplitude $A$ represents a series of parton emissions 
from an initial parton with momentum $k$ which emerges with a momentum $l$ and then undergoes another series 
of splittings represented by the kernel B. 
The spin projectors for the collinear case can be summarised, for the gluon and quark case respectively, as
\bea
\Pbb_{g,\,\text{in}}^{s\,\mu\nu} 
&=&
\frac{1}{d-2}\left(- g^{\mu\nu} + \frac{l^\mu n^\nu + n^\mu l^\nu}{l\cdot n} \right) \, , 
\quad
\Pbb_{g,\,\text{out}}^{s\,\mu\nu}  = - g^{\mu\nu} \, ,  
\nn \\
\Pbb^{s}_{q,\,\text{in}} 
&=& 
\frac{ \slashed{l}}{2} \, ,
\quad \quad \quad \quad \quad \quad \quad \quad \quad \quad \quad \quad \quad \, \, \, \,  
\Pbb^{s}_{q,\,\text{out}} = \frac{\slashed{n}}{2\,n\cdot l}\, ,
\label{coll_proj}
\eea
where we have omitted the spinor indices for the sake of brevity, as we will be doing through the rest of this paper.
Two comments are in order here:
\begin{enumerate}
\item 
The incoming projectors yield the  average over the gluon or quark helicities in $d$ dimensions.
\item 
The projectors are not uniquely defined: 
both the momentum projector $\Pbb^{\epsilon}$ (due to the
arbitrariness of the factorization scheme) and the spin projectors are
defined modulo finite terms and a re-definition is possible as long as
all singular terms are properly extracted.
\end{enumerate}
For the sake of completeness,  we further report here the action of the whole projector on the product of two kernels
in the case of an intermediate quark state,
in order to clarify the role of the momentum space projector $\Pbb^\epsilon$;
it is explicitly given by
\bea
A \, \Pbb_C \,B 
&\equiv&
A \, \Pbb^\epsilon \otimes \Pbb^s \,B = 
A(q, l)_{\dots\alpha\alpha'}\, \frac{(\slashl)^{\alpha\alpha'}}{2}\, \Pbb^{\epsilon}\, \int \frac{d^ml}{(2\pi)^m}\, \frac{(\slashn)_{\beta\beta'}}{2n\cdot l}\, B^{\dots\beta\beta'}(l,k) \, ,
\label{full_proj}
\eea
where $\Pbb^\epsilon$ sets $l^2 = 0 $ in the $A$ kernel and
takes the pole part (or pole part plus finite terms, depending on the scheme choice) of the integral
over $d^d l \propto dl^2\, d^{d-2}l_\perp $, with the integral over $l^2$ defined up to the factorization scale.%
    \footnote{This clearly has to be the case, but since the projector selects only the infrared poles,
    which come anyway from the region of low $l^2$, the upper limit of the integral is actually irrelevant.
    However, starting from NLO more complicated forms of the upper phase space limit can lead to the dependence
    of the splitting functions on this choice~\cite{Jadach:2016zgk}.}
The reason for setting $l^2 = 0$ in the left-hand kernel is clear in light of the proof of~\cite{Ellis:1978sf} that the
infrared singularities for a cross section which is 2PI expanded in light-like gauge come only from the integral over the intermediate 
propagator momenta, which, by definition, are attached only to the outgoing lines of the kernels.

\subsection{Generalization to the TMD case}

In the  case where the incoming particle has a transverse momentum component parametrised
as in (\ref{eq:kinematics}), an economic choice for the projectors has been
introduced by Catani and Hautmann in~\cite{Catani:1994sq} (gluon case)  and two of us in~\cite{Gituliar:2015agu}  (quark case):  
\begin{eqnarray}
l = y\, p + l_\perp\, , \quad \Rightarrow
&&
\left\{ \begin{array}{lc}
\Pbb_{g,\,\text{in}}^{s\,\mu\nu} = -\frac{l_\perp^\mu\, l_\perp^\nu}{l_\perp^2} & \quad \Pbb_{g,\,\text{out}}^{s\,\mu\nu}  = - g^{\mu\nu} 
\\
\Pbb^{s}_{q,\,\text{in}} = \frac{ y\,\slashp}{2} & \; \Pbb^{s}_{q,\,\text{out}} = \frac{\slashed{n}}{2\,n\cdot l} \\
\end{array} \right.\,,
\label{CH_proj}
\end{eqnarray}
where the difference to the collinear case Eq.~(\ref{CH_proj}) lies in the incoming projectors.
While for the quark it is easy to see that the numerator of the incoming projector merely provides the 
collinear part of the momentum dotted into the gamma matrices 
(and actually coincides completely with the collinear case), for the gluon  both projectors are at first considerably different. 
While such polarizations for high energy factorized gluons are well
known in the literature and can be traced back to a gauge rotation of
the original `non-sense' polarization $\sim p^\mu, n^\mu$ of a
reggeized gluon (see \cite{Forshaw:1997dc} for a pedagogic
introduction) it is interesting to take a closer look at the precise
motivation given by the authors of~\cite{Catani:1994sq}, which
essentially refers to~\cite{Catani:1990eg}. 
In a nutshell,~\cite{Catani:1990eg} considers the amplitude for heavy quark
pair production via the fusion of two off-shell gluons. As outlined in
the Appendix B of~\cite{Catani:1990eg}, this amplitude is defined to
contain the polarization tensor in axial gauge for each of
two off-shell gluons.  The open gluon indices are then contracted by
`non-sense' polarization vectors $\sim p^\mu$ and $\sim n^\mu$, 
as it is adequate for high energy factorized gluons. The crucial observation 
is then, that contraction of the polarization tensor in axial gauge
with `non-sense' polarization vectors provides a vector
proportional to the transverse momentum of the regarding gluon. In our
case this relation is encoded in the identity
\begin{align}
 \label{CHtrick} 
   y\, p^\mu\, d_{\mu\nu}(k) & = k_{\perp\,\nu}  & \text{if} &&   k^\mu = y\,p^{\mu} + k^\mu_\perp\, .
\end{align}
In turn, it is then possible to simply work with external off-shell
gluonic legs with the axial gauge tensor of the external gluonic legs already stripped off. 
However, in our case the scenario is slightly different: 
since the generalized production vertex requires the presence of the polarization tensors 
in axial gauge and these cannot be factorized, we must simply contract the open index of our initial high
energy factorized gluon with $y p^\mu$, as suggested by the original formulation of high energy factorization,
with no possibility for further simplifications. 
However, doing so seems to lose one of initial advantage of the choice
$k_\perp^\mu k_\perp^\nu / k_\perp^2$, i.e. that, using the same "out"
projector as for the collinear case,  the resulting modified projector
$$
\Pbb^s_g =  -\frac{k_\perp^{\mu'} k_\perp^{\nu'}}{k_\perp^2} \, (-g_{\mu\nu}) \, ,
$$
satisfies the necessary constraints of being equal to its square 
and reducing to the collinear projector after integration over 
the azimuthal angle, in the limit of vanishing transverse momentum
\bea
\Pbb^s_{g} \otimes \Pbb^s_{g} 
&=&  
\frac{k_\perp^{\mu'}\, k_\perp^{\nu'}}{k_\perp^2} \, (-g_{\mu\nu}) \, \frac{k_\perp^{\mu}\, k_\perp^{\nu}}{k_\perp^2} \, (-g_{\rho\sigma}) 
= -\frac{k_\perp^{\mu'}\, k_\perp^{\nu'}}{k_\perp^2} \, (-g_{\rho\sigma})= \Pbb^s_{g} \, ,
\label{solid1} 
\\
\Big\langle \frac{k_\perp^{\mu}\, k_\perp^{\nu}}{k_\perp^2} \Big\rangle_\phi
&\stackrel{k_\perp \rightarrow 0}{=}& 
\frac{d^{\mu\nu}(k = z\,p)}{d-2} \, .
\label{solid2}
\eea 
Since for the 3-gluon vertex $\Gamma^{\mu_1\mu_2\mu_3}_{g_1^*g_2^* g_3}$
it is not possible to factor out the numerators of the gluon propagators,
the choice of the Catani-Hautmann projector  is unfeasible for our purposes,
which makes 
\beq
\Pbb^s_{g\, \text{in}} =  -y^2\, \frac{p^{\mu} p^{\nu}}{k_\perp^2}
\label{ourin}
\eeq
the natural choice for the  incoming projector. 
However, in order for (\ref{solid1}) to hold, we cannot keep the same "out" projector. 
It is nevertheless possible to use the projector proposed already in \cite{Gituliar:2015agu},
\beq
\Pbb^s_{g\, \text{out}} = -g^{\mu\nu} + \frac{k^\mu n^\nu + k^\nu n^\mu }{k\cdot n} - k^2\, \frac{n_\mu n_\nu}{(k\cdot n)^2}\, .
\label{modout}
\eeq
It is easy to check that this restores the condition (\ref{solid1}),
$\Pbb^s\otimes\Pbb^s = \Pbb^s$. Moreover, as already noted
in~\cite{Gituliar:2015agu} the new "out" projector, by itself, is
consistent with the collinear case. In an actual calculation all terms
apart from the $-g_{\mu\nu}$ are set to zero by the polarization
tensor of the gluon propagators connecting to the projector, due to
the property $d_{\mu\nu} \cdot n^\nu = 0 = d_{\mu\nu} \cdot n^\mu$.
We explicitly checked that changing the "out" projector in
the collinear case does not modify the LO collinear splitting functions. 

What remains to be checked is that the collinear limit holds also for
the "in" projector. Since the numerator is purely longitudinal, there
exists no relation similar to Eq.~(\ref{solid2}) after angular averaging, 
at least not at the general operator level.  
However, it can be proved that the difference between
$y^2\, p^\mu p^\nu/k_\perp^2$ and $k^\mu_\perp k^\nu_\perp/k_\perp^2$
vanishes when they are contracted into the relevant vertices.

We have to show this for two vertices,
$\Gamma^\mu_{q^*g^*q}(q,k,p')$, (for the $\tilde{P}_{qg}$ kernel), and
$\Gamma^{\mu_1\mu_2\mu_3}_{g^*g^*g}(k,q,p')$, (for $\tilde{P}_{gg}$). 
		
The first case is trivial since due to the factor $d^{\mu\nu}(k)$,  the projector
$y^2\, p^\mu p^\nu / k_\perp^2$ automatically reduces to
$k_\perp^\mu k_\perp^\nu / k_\perp^2$ due to  Eq.~(\ref{CHtrick}). 

As for $\Gamma^{\mu_1\mu_2\mu_3}_{g^*g^*g}(k,q,p')$, it is easy to check that
\beq
k_{\mu_1}\, \Gamma^{\mu_1\mu_2\mu_3}_{g^*g^*g}(k,q,p')
= \mathcal{O}(k_\perp^2) \, .
\eeq
As a consequence we find
\begin{align}
  \label{eq:1}
y\, p_{\mu_1}\, \Gamma^{\mu_1\mu_2\mu_3}_{g^*g^*g}(k,q,p') = 
- k_{\perp\, \mu_1}\,  \Gamma^{\mu_1\mu_2\mu_3}_{g^*g^*g}(k,q,p') +  \mathcal{O}(k_\perp^2) \, ,
\end{align}
which is sufficient to establish agreement in the collinear limit,
at least at the perturbative order which we are considering; 
indeed, this means that the collinear limit is exactly the same as for the Catani-Hautmann projector.
Therefore, the final set of projectors which we will use in the following is  given by
\bea
\Pbb_{g,\,\text{in}}^{s\,\mu\nu} 
&=&
-y^2\, \frac{p^{\mu} p^{\nu}}{k_\perp^2} \, , 
\quad\quad
\Pbb_{g,\,\text{out}}^{s\,\mu\nu}  = 
-g^{\mu\nu} + \frac{k^\mu n^\nu + k^\nu n^\mu }{k\cdot n} - k^2\, \frac{n_\mu n_\nu}{(k\cdot n)^2} \, ,  
\nn \\
\Pbb^{s}_{q,\,\text{in}} 
&=& 
\frac{ y\,\slashp}{2} \, ,
\quad \quad \quad \quad \, \, \, \,  
\Pbb^{s}_{q,\,\text{out}} = \frac{\slashed{n}}{2\,n\cdot l}\, .
\label{eq:HE_proj_new}
\eea
Finally, let us note that the convolution product we have just dissected in detail is precisely the one 
in Eq.~(\ref{eq:TMDkernelDefINI}) from which we derived the definition of our TMD splitting functions.
For the collinear case, an all-order argument for the derivation of splitting functions is presented 
in~\cite{Curci:1980uw}, to which we refer the interested reader.

\subsection{Gauge invariance of the effective production vertex}
\label{sec:gauge-invar-effect}
The effective gluon production vertex,
\begin{align}
\label{eq:ggg_vertexX} 
\Gamma^{\mu_1\mu_2\mu_3}_{g^*g^*g}(q,k,p') &  =
 \mathcal{V}^{\lambda \kappa \mu_3}(-q,k,-p') \, {d^{\mu_1}}_{\lambda} (q)\, {d^{\mu_2}}_{\kappa}(k)  \notag \\
&  \hspace{2cm}
+ d^{\mu_1\mu_2}(k)\, \frac{q^2 n^{\mu_3}}{ n\cdot p'} - d^{\mu_1\mu_2}(q)\, \frac{k^2 p^{\mu_3}}{ p\cdot p'}
\end{align}
obtained in Eq.~\eqref{eq:ggg_vertex} is so far still restricted to
pure high energy kinematics, where $n \cdot q = 0$ and therefore
$n \cdot k = n \cdot p' $. This no longer holds for the more general TMD
kinematics Eq.~\eqref{eq:kinematics}. While the QCD three gluon vertex
is already fixed for general momenta, $n \cdot k \neq n \cdot p' $
leaves us at first with an ambiguity for the denominator of the second
term in Eq.~\eqref{eq:ggg_vertex} which cannot be fixed by high energy
factorization alone. Similar to the case of the quark splitting
functions, we find that this ambiguity can be solved if we require
current conservation for the produced real gluon. To this end we first recall the Ward identity of the QCD three gluon vertex in the case where both gluons are off-shell, 
\begin{align}
  \label{eq:2}
  \mathcal{V}^{\lambda \kappa \mu_3}(-q,k,-p') \cdot p'_{\mu_3} & = \left(k^{\lambda} k^{\kappa} - q^\lambda q^\kappa \right) + \left(q^2 - k^2 \right)g^{\lambda\kappa}\, .
\end{align}
Adding polarization tensors and contraction with the
polarization of the incoming (high energy factorized) gluon, $y p_{\mu_2}$, one has
\begin{align}
  \label{eq:6}
   \mathcal{V}^{\lambda \kappa \mu_3}(-q,k,-p') \cdot p'_{\mu_3} \cdot d^{\mu_1}{}_{\lambda}(q) d^{\mu_2}{}_\kappa y p_{\mu_2} & =  k^2 ( d^{\mu_1}{}_{\lambda}(q ) y p^\lambda) - q^2  ( d^{\mu_1}{}_{\lambda}(k ) y p^\lambda).
\end{align}
On the other hand one finds for the sum of non-local terms
\beq
\label{eq:4}
y\,p_{\mu_2} \left(    d^{\mu_1\mu_2}(k)\, \frac{q^2 n^{\mu_3}}{ n\cdot p'} - d^{\mu_1\mu_2}(q)\, \frac{k^2 p^{\mu_3}}{ p\cdot p'} \right)  \cdot p_{\mu_3}'   =
y\,p_{\mu_2} \bigg( d^{\mu_1\mu_2}(k)\, q^2  - d^{\mu_1\mu_2}(q)\,  k^2 \bigg) \, ,
\eeq
which cancels precisely the terms arising in Eq.~\eqref{eq:6} and we
demonstrated current conservation with respect to the produced gluon.
Note that this is only achieved if we fix the denominator of the second
term in Eq.~\eqref{eq:ggg_vertexX} to the form given above. 

\subsection{Comparison to  Lipatov's effective action}
\label{sec:comp-ggr-vert}

At this stage we finally return to the discussion at the beginning of
Sec.~\ref{vertices}. Calculating the gluon-gluon-reggeized gluon (GRR)
vertex in $A \cdot n = 0$ light-cone gauge from Lipatov's high energy
effective action,  where the reggeized gluon
is identified with the incoming gluon with momentum $k$ and the gluons
with momenta $q$ and $p'$ are treated as conventional QCD gluons, one finds
\begin{align}
  \label{eq:13}
 \Gamma_{GGR}^{\lambda\mu_3}(q,k,p') & \propto p_{\mu_2} d^{\mu_2}_{\kappa}(k)  \mathcal{V}^{\lambda
  \kappa \lambda}(-q,k,-p') \,
 -
  p^{\lambda} p^{\mu_3}\frac{k^2}{ p\cdot p'} \, ,
\end{align}
where `$\propto$' merely serves to indicate that the above expression
does not coincide with the overall normalization as provided within
Lipatov's effective action.  Adding the polarization tensor of the
gluon with momentum $q$ to this vertex,  one  finds easily
\begin{align}
  \label{eq:14}
  d^{\mu_1}{}_\lambda(q) \Gamma_{GGR}^{\lambda\mu_3}(q,k,p') & \propto   p_{\mu_2} 
\bigg[ 
  \mathcal{V}^{\lambda\kappa \lambda}(-q,k,-p')   d^{\mu_2}_{\kappa}(k)  
d^{\mu_1}{}_\lambda(q)
 -
 d^{\mu_2}{}_\lambda(q) \frac{k^2 p^{\mu_3}}{p \cdot p'}
 \bigg].
\end{align}
Since within light-cone gauge the term proportional to $n^{\mu_3}$ in
Eq.~\eqref{eq:ggg_vertexX} will be set to zero, use of the above
expression is equivalent to the use of Eq.~\eqref{eq:ggg_vertexX}
within the employed modification of the CFP formalism. However, 
due to the absence of the term proportional to $n^{\mu_3}$, 
a direct verification of gauge invariance is not possible.

\section{Results for  the real emission splitting functions}
\label{sec:results}

In this section we present our results for the  TMD splitting functions. 
When performing the calculations, we follow exactly the steps that are outlined in Sec.~\ref{real}, 
using the modified vertices and projectors defined in Secs.~\ref{vertices} and~\ref{sec:proj}.
We discuss in more detail the case of the $\tilde{P}_{gg}$ kernel, as this is the first time this TMD splitting
functions is being calculated using a CFP inspired framework.

\subsection{Quark splitting functions}
\label{sec:quark-splitt-funct}

For the quark splitting functions previously computed in \cite{Catani:1994sq,Hautmann:2012sh,Gituliar:2015agu} 
we confirm the previous results, after including the modification discussed in the foregoing section. 
For completeness we present here the precise expressions for the TMD splitting functions
\begin{eqnarray}
\label{eq:5}
\tilde{P}_{qg}^{(0)} 
&=& 
T_R\, \left(\frac{\qtt^2}{\qtt^2+z(1-z)\kt^2}\right)^2 \nn \\
&\times&
\left[ 1 + 4z^2(1-z)^2\frac{\kt^2}{\qtt^2} + 4z(1-z)(1-2z)\frac{\kt\cdot\qtt}{\qtt^2} - 4z(1-z)\frac{(\kt\cdot\qtt)^2}{\kt^2\qtt^2} \right] ,\\
%
\tilde{P}_{gq}^{(0)} 
&=& 
C_F\, 
\left(\frac{\qtt^2}{\qtt^2+z(1-z)\kt^2}\right)^2\,  \frac{\qtt^2}{(\qtt-(1-z)\kt)^2} 
\nn \\
&\times& 
\bigg[ \frac{2}{z} - 2 + z + 2(1-z)(1+z-z^2)\frac{\kt^2}{\qtt^2} + z(1-z)^2(1+z^2)\frac{\kt^4}{\qtt^4}
\nn \\
&+&
4z^2(1-z)^2\frac{\kt^2 \, \kt\cdot\qtt}{\qtt^4} + 4(1-z)^2\frac{\kt\cdot\qtt}{\qtt^2} + 4z(1-z)^2\frac{(\kt\cdot\qtt)^2}{\qtt^4} \bigg]    
\nn \\
&+& \epsilon\, C_F\, \frac{z\qtt^2 \left(\qtt-(1-z)\kt\right)^2}{(\qtt^2+z(1-z)\kt^2)^2} \, ,
\\
%
\tilde{P}_{qq}^{(0)} 
&=&
C_F\, 
\left(\frac{\qtt^2}{\qtt^2+z(1-z)\kt^2}\right)^2 \frac{\qtt^2}{(\qtt-(1-z)\kt)^2}
\nn \\
&\times&
\bigg[  \frac{1+z^2}{1-z} + (1+z+4z^2-2z^3)\frac{\kt^2}{\qtt^2} + z^2(1-z)(5-4z+z^2)\frac{\kt^4}{\qtt^4}
\nn \\
&+& 
2z(1-2z)\frac{\kt\cdot\qtt}{\qtt^2} + 2z(1-z)(1-2z)(2-z)\frac{\kt^2\, \kt\cdot\qtt}{\qtt^4}
\nn \\
&-&  4z(1-z)^2\frac{(\kt\cdot\qtt)^2}{\qtt^4}
\bigg]  + 
\epsilon\, C_F\, \frac{(1-z)\qtt^2(\qtt+z\kt)^2}{(\qtt^2+z(1-z)\kt^2)^2} \, ,
\end{eqnarray}
and for the angular averaged TMD splitting functions (with $\epsilon=0$)
\begin{eqnarray}
\label{eq:5}
\bar{P}_{qg}^{(0)} 
&=&
T_R \left(\frac{\qtt^2}{\qtt^2+z(1-z)\kt^2}\right)^2\, 
\left[ z^2 + (1-z)^2 + 4z^2(1-z)^2\frac{\kt^2}{\qtt^2}
\right],
\\
%
\bar{P}_{gq}^{(0)} 
&=& 
C_F\, \bigg[
\frac{2\qtt^2}{z|\qtt^2-(1-z)^2\kt^2|}
- \frac{(2-z)\qtt^4+z(1-z^2)\kt^2\qtt^2}{\left(\qtt^2+z(1-z)\kt^2\right)^2}
\bigg] \, ,
\\
%
\bar{P}_{qq}^{(0)} 
&=&
C_F\,  \frac{\qtt^2}{\qtt^2+z(1-z)\kt^2}
\nn \\
&\times&
\bigg[  
\frac{\qtt^2+(1-z^2)\kt^2}{(1-z)|\qtt^2-(1-z)^2\kt^2|} + \frac{z^2\qtt^2-z(1-z)(1-3z+z^2)\kt^2}{(1-z)(\qtt^2+z(1-z)\kt^2)} 
\bigg] \, .
\end{eqnarray}
It is easy to check that the above splitting functions reduce to the
collinear DGLAP results when $\kt^2\to0$.

\subsection{The gluon-to-gluon splitting function}

The real part of the $\tilde{P}_{gg}$ splitting function is given by the matrix element originating from the diagram in Fig.~\ref{fig:pgg}
\bea
&&
g^2\, \delta\left((k-q)^2\right)\, W_{gg} =
\Pbb_{g,\,\text{in}} \otimes \hat{K}_{gg}^{(0)}(q, k) \otimes \Pbb_{g,\,\text{out}} =
\nn \\
&&
\Pbb_{g,\,\text{in}}^{\beta\beta'}(k) \, \mathbb{P}_{g,\,\text{out}}^{\mu'\nu'}(q)
(\Gamma_{g^*g^*g}^{\beta\mu\alpha})^\dagger
\Gamma_{g^*g^*g}^{\nu\beta'\alpha'} \,
\frac{-i d^{\mu\mu'}(q)}{q^2 - i\epsilon} \,
\frac{id^{\nu\nu'}(q)}{q^2 + i \epsilon}
 \,d^{\alpha\alpha'}(k-q),
\eea
where $\Gamma_{g^*g^*g}^{\mu\nu\alpha}$ is the effective gluon
production vertex of Eq.~\eqref{eq:ggg_vertexX} and we use the newly
defined gluon projector Eq.~\eqref{eq:HE_proj_new}.  We obtain the
following $\tilde{P}_{gg}$ splitting function
\begin{align}
  \label{eq:ggsplitting}
  \tilde{P}_{gg}^{(0)} (z, \qtt, \kt)
 &=
 2 C_A\, \bigg\{
 \frac{\qtt^4}{\left(\qtt-(1-z)\kt\right)^2[{\qtt^2+z(1-z)\kt^2}]} 
\bigg[\frac{z}{1-z} + \frac{1-z}{z}  +
\notag \\
& \hspace{-1.5cm}  + 
(3-4z) \frac{\qtt \cdot \kt}{ \qtt^2} + z(3-2z) \frac{\kt^2}{\qtt^2}
\bigg] + \frac{(1 + \epsilon)\qtt^2 z(1-z) [2 \qtt \cdot \kt + (2z -1)
  \kt^2]^2}{2 \kt^2 [\qtt^2+z(1-z)\kt^2]^2} \bigg\} \, .
\end{align}
After angular averaging (and setting $\epsilon=0$) this provides
\bea
\bar{P}_{gg}^{(0)}\left(z, \frac{\kt^2}{\qtt^2} \right) 
&=&
C_A\, 
\frac{\qtt^2}{\qtt^2+z(1-z)\kt^2}
\bigg[
\frac{(2-z)\qtt^2+(z^3-4z^2+3z)\kt^2}{z(1-z)\left|\qtt^2-(1-z)^2\kt^2\right|}
\nn \\
&+& 
   \frac{(2z^3-4z^2+6z-3)\qtt^2+z(4z^4-12z^3+9z^2+z-2)\kt^2}{(1-z)(\qtt^2+z(1-z)\kt^2)}
\bigg] \, .
\eea
%

\subsection{Kinematic limits}
\label{sec:kinematic-limits}

As a next step we verify the necessary kinematic limits which the kernel needs to obey. In the collinear limit this is straightforward, 
since the transverse integral in Eq.~\eqref{eq:TMDkernelDefINI} is specially adapted for this limit. 
In particular, one easily obtains the real part of the DGLAP gluon-to-gluon splitting 
function:\footnote{We verified this limit also for finite $\epsilon$ where it holds equally.}
\begin{align}
\label{eq:7}
\lim_{\kt^2 \to 0} \bar{P}_{ij}^{(0)}\left(z, \frac{\kt^2}{\qtt^2}\right) &= 2\, C_A\, \left[ \frac{z}{1-z} + \frac{1-z}{z} + z\,\left(1-z\right) \right] \, .
\end{align}
In order to study the behaviour of the obtained splitting kernel in the high
energy and soft limit, it is useful to change the variables of
integrations in the TMD kernel Eq.~\eqref{eq:TMDkernelDefINI}
which will be particularly useful to disentangle $z\to 1$ and the ${\qtt \to  (1-z) \kt }$ singularities. 
With $\ptt = \frac{\kt - \qt}{1-z} = \kt - \frac{\qtt}{1-z}$ and changing variables accordingly we have
\begin{align}
\label{eq:TMDkernelDefptilde}
\hat K_{gg} \left(z, \frac{\kt^2}{\mu^2}, \epsilon, \alpha_s \right) 
 &=
C_A\, \frac{\alpha_s}{2\pi} \, z\,  
\frac{e^{-\epsilon\gamma_E}}{\mu^{2\epsilon}}
\int \frac{d^{2+2\epsilon}\ptt}{ \pi^{1+\epsilon}} \,  
\Theta\left(\mu_F^2  - ( (1-z) (\kt - \ptt)^2 + z \kt^2)\right) \, 
\notag \\
& \times \hspace{0cm} (1-z)^{2\epsilon} \bigg[\frac{2}{(1-z)z\, \ptt^2}
 + 
   \frac{2 (\ptt \cdot \kt - 2 (1-z) \ptt^2 )}{\ptt^2  [ (1-z) (\kt -
  \ptt)^2 + z \kt^2 ] }  
\notag \\
& \hspace{2cm} 
+
(1+ \epsilon)\,  \frac{z(1-z) }{ \kt^2 } 
 \left( 
\frac{\kt^2 - 2 (1-z) \ptt \cdot \kt}{ (1-z) (\kt - \ptt)^2 + z \kt^2} \right)^2
\bigg]
\end{align}
where 
\begin{align}
  \label{eq:9}
 - q^2 &=  (1-z) (\kt - \ptt)^2 + z \kt^2\,,
\end{align}
yields precisely the absolute value of the virtuality of the $t$-channel gluon.
First we note that in the collinear limit $\ptt^2 \gg \kt^2$ we
re-obtain the DGLAP splitting function, Eq.~\eqref{eq:7}, also in this
parametrization. In the high energy limit $z \to 0$ we obtain
\begin{align}
  \label{eq:8}
  \lim_{z\to 0} \hat K_{gg} \left(z, \frac{\kt^2}{\mu^2}, \epsilon, \alpha_s \right)  
&=
\frac{\alpha_s C_A}{\pi (e^{\gamma_E}\mu^2)^\epsilon}\int \frac{d^{2 + 2 \epsilon} \ptt}{\pi^{1 + \epsilon}} \Theta\left(\mu_F^2 - (\kt - \ptt)^2\right) \frac{1}{\ptt^2} \notag \\
& = 
\int \frac{d^{2 + 2 \epsilon} \qt}{\pi^{1 + \epsilon}} \Theta\left(\mu_F^2 -{\qt}^2\right)  \frac{\alpha_s C_A}{\pi (e^{\gamma_E}\mu^2)^\epsilon} \frac{1}{(\qt - \kt)^2},
\end{align}
where the term under the integral is easily identified as the real
part of the LO BFKL kernel. For the
limit $z \to 1$,  we find, in complete analogy,
\begin{align}
  \label{eq:10}
  \lim_{z\to 1} \hat K_{gg} \left(z, \frac{\kt^2}{\mu^2}, \epsilon, \alpha_s \right)  
&=
\frac{\alpha_s C_A}{\pi (e^{\gamma_E}\mu^2)^\epsilon}\int \frac{d^{2 + 2 \epsilon} \ptt}{\pi^{1 + \epsilon}} \Theta\left(\mu_F^2 - \kt^2\right) \frac{1}{\ptt^2 (1-z)^{1-2\epsilon}} \notag \\
&  \hspace{-3cm}=
\frac{\alpha_s C_A}{\pi (e^{\gamma_E}\mu^2)^\epsilon}\int \frac{d^{2 + 2 \epsilon} \ptt}{\pi^{1 + \epsilon}} \Theta\left(\mu_F^2 - \kt^2\right) \frac{1}{\ptt^2} \left( \frac{1}{2 \epsilon} \delta(1-z) + \frac{1}{(1-z)_+^{1-2\epsilon}}\right),
\end{align}
where in the last term we made use of the plus prescription to isolate
the singularity at $z=1$ (see \cite{Gituliar:2015agu} or any QCD
standard reference for the precise definition).\\

  Another limit of
interest is the vanishing transverse momentum of the produced gluon
$\ptt \to 0$.  If we parametrize the phase space of the produced gluon
in terms of its transverse momentum, rapidity and azimuthal angle,
this limit corresponds to the soft limit, {\it i.e.} the limit where
the four momentum of the produced gluon vanishes.  To isolate this
singularity, we introduce a phase space slicing parameter $\lambda$,
to separate the potential divergent region ($|\ptt| < \lambda$) from
the finite transverse momentum integral ($|\ptt| > \lambda$).  Taking
the limit $\lambda \to 0$ we find for the divergent part of the TMD
kernel,
\begin{align}
\label{eq:11}
\hat K_{gg}^{\text{div.}} \left(z, \frac{\kt^2}{\mu^2}, \epsilon, \alpha_s \right)  
&=\Theta(\mu_F^2 - \kt^2) \frac{\alpha_s C_A}{\pi} \frac{e^{-\gamma_E \epsilon}}{\epsilon \Gamma(1 + \epsilon)} \left(\frac{\lambda^2}{\mu^2} \right)^\epsilon \frac{1}{(1-z)^{1-2\epsilon} z} \notag \\
& \hspace{-2cm}=\Theta(\mu_F^2 - \kt^2) \frac{\alpha_s C_A}{\pi} \frac{e^{-\gamma_E \epsilon}}{\epsilon \Gamma(1 + \epsilon)} 
\left(\frac{\lambda^2}{\mu^2} \right)^\epsilon \bigg[\frac{1}{2\epsilon} \delta(1-z) +  \frac{1}{(1-z)_+^{1-2\epsilon} z} \bigg].
\end{align}
As usually the $\ptt \to 0$ and the $z\to 1$ singularity are
regularized by dimensional regularization, while the high energy
singularity requires a separate regulator (or needs to be controlled
by energy conservation).  Apart from the above extraction of
singularities, it is also interesting to simply consider the
$\ptt \to 0$ limit of the TMD kernel, where set for the time being
$\epsilon \to 0$; one obtains
\begin{align}
\label{eq:12}
\hat K_{gg} \left(z, \frac{\kt^2}{\mu^2}, 0, \alpha_s \right)  & =
z\,  \int_{\ptt^2_{min}}^{\ptt^2_{max}} \frac{d \ptt^2}{\ptt^2} \frac{\alpha_s C_a}{\pi}\bigg[\frac{1}{z} + \frac{1}{1-z} + \mathcal{O}\left(\frac{\ptt^2}{\kt^2}\right) \bigg] \, ,
\end{align}
which allows to identify the terms under the integrand as the (real
part/unresummed\footnote{Both Sudakov and Non-Sudakov-form factors are
  obviously absent and the lower and upper limits of integration are
  result of adding virtual corrections and appropriate angle related
  scale of hard process.}) CCFM-kernel.  Note that we did not take any
dedicated measures to arrive at this limit; it merely appears as a
natural by-product of the conditions imposed on the TMD splitting
kernel.

\section{Summary and discussion}
\label{sec:conclusions}
The main result of this paper is the calculation of a transverse
momentum dependent gluon-to-gluon splitting function.  The splitting
function reduces both to the conventional gluon-to-gluon DGLAP
splitting function in the collinear limit as well as to the LO BFKL
kernel in the low $x$/high energy limit; moreover the CCFM
gluon-to-gluon splitting function is re-obtained in the limit where
the transverse momentum of the emitted gluon vanishes, {\it i.e.} if
the emitted gluon is soft.  The derivation of this result is based on
the Curci-Furmanski-Petronzio formalism for the calculation of DGLAP
splitting functions in axial gauges.  To address gauge invariance in
the presence of off-shell partons, high energy factorization adapted
for axial gauges has been used to derive an effective production
vertex which then could be shown to satisfy current conservation.
\\

The next step in completing the calculation of TMD splitting functions
is the determination of the still missing virtual corrections, which
will be carried out using the techniques developed in
\cite{Hentschinski:2011tz,Chachamis:2012cc,Chachamis:2012mw,Hentschinski:2011xg,Chachamis:2012gh,Chachamis:2013hma}
within Lipatov's high energy effective action and/or the helicity
spinor framework~\cite{vanHameren:2017hxx}.  With the complete set of
splitting functions at hand, it will be finally possible to formulate
an evolution equation for the unintegrated (TMD) parton distribution
functions including both gluons and quarks.  As a final goal, we have
in mind to use this evolution equation to formulate a parton shower
algorithm, which can be implemented into a new Monte Carlo program
with TMD splitting functions, extending currently available codes such
as \texttt{Cascade} \cite{Jung:2010si}.  This will allow to address,
with new theoretical tools within the TMD approach,
both low $x$ and especially moderate $x$ phenomenology \cite{Motyka:2016lta}.\\

Another direction for future research will aim at clarifying the
relation of our result with the linearised results for TMD evolution
in the low and intermediate $x$ region obtained
in~\cite{Balitsky:2015qba}. While both results agree in the collinear
and the high energy limit, they differ for more general kinematics.
We believe that the origin of this difference lies in an observation
made in \cite{Dominguez:2011wm}, see also \cite{Kotko:2015ura}, namely
the existence of 2 different gluon distributions in the low $x$
limit. While the TMD distribution studied in~\cite{Balitsky:2015qba}
-- which agrees in the intermediate $x$ region with the definition
provided in \cite{Mulders:2000sh} -- seems to correspond to the
Weizs\"acker-Williams gluon distribution (in the terminology
of~\cite{Dominguez:2011wm}), the gluon distribution obtained from
``unintegrating'' the collinear gluon distribution underlying the
CFP-formalism, appears at least at first sight related to what
\cite{Dominguez:2011wm} call the Fourier-transform of the dipole
distribution; in particular it is said dipole distribution which
enters inclusive observables in the high energy limit {\it e.g.}
proton structure functions.  A detailed answer and study of this
question is beyond the scope of this paper and will be addressed in a
separate publication.  In particular, such a research also needs to
answer the question as to which extent the obtained TMD kernels can be
derived independently from the use of the Curci-Furmanski-Petronzio
formalism, which is constrained to the use of axial light-cone gauges.

\section*{Acknowledgments}
We would like to thank A. van Hameren for useful discussions. 
The work of MS and KK is supported by the NCN grant DEC-2013/10/E/ST2/00656. 
MS is also partially supported by the  Israeli Science Foundation through grant 1635/16, 
by the BSF grants 2012124 and 2014707, by the COST Action CA15213 THOR
and by a Kreitman fellowship by the Ben Gurion University of the Negev. 
MS and AK wish to thank the \emph{Laboratoire de Physique Subatomique et de Cosmologie}
in Grenoble, particularly its Particle Theory Group, for the warm
hospitality during the initial stage of this project. MH is grateful for
the warm hospitality with which he was received at the H. Niewodnicza\'nski
Institute of Nuclear Physics and partial support from Grant of
National Science Center, Poland, No. 2015/17/B/ST2/01838.
All figures were drawn with \emph{feynMF}~\cite{Ohl:1995kr}.


\appendix

\section{Conventions for the spinor helicity amplitudes}
\label{AppSpinors}%

We repeat here the parametrisation of the momenta used 
in~\cite{vanHameren:2014iua,vanHameren:2015bba}, which reads
\begin{align}
k_1^\mu + k_2^\mu + \cdots + k_n^\mu = 0
&\qquad\textrm{momentum conservation}\label{Eq:momcons}\\
p_1^2 = p_2^2 = \cdots = p_n^2 = 0
&\qquad\textrm{light-likeness}\label{momcon1}\\
\sp{p_1}{k_1} = \sp{p_2}{k_2}=\cdots=\sp{p_n}{k_n}=0
&\qquad\textrm{eikonal condition}
\label{Eq:eikcon}
\end{align}
where for every $k^\mu_i$ there is a corresponding orthogonal, light-like direction $p^\mu_i$.

With the help of an auxiliary light-like four-vector $q^\mu$, the momentum $k^\mu$ 
can be decomposed in terms of its light-like direction $p^\mu$, satisfying $\sp{p}{k}=0$, and a transversal part, following
\begin{equation}
k^\mu = x(q)p^\mu 
- \frac{\kapp}{2}\,\frac{\AS{p|\gamma^\mu|q}}{\SSS{pq}} 
- \frac{\kstr}{2}\,\frac{\AS{q|\gamma^\mu|p}}{\AA{qp}} \, ,
\end{equation}
with
\begin{equation}
x(q)=\frac{\sp{q}{k}}{\sp{q}{p}}
\quad,\quad
\kapp = \frac{\AS{q|\slashK|p}}{\AA{qp}}
\quad,\quad
\kstr = \frac{\AS{p|\slashK|q}}{\SSS{pq}} \, .
\end{equation}
If the momentum $k$ is on-shell, then it coincides with its associated direction, $p=k$.
\begin{equation}
k^2 = -\kapp\kstr \ .
\end{equation}

We focus on the fundamental {\em colour-ordered\/} or {\em dual\/} amplitudes,
where the gauge-group factors have been stripped off and all particles are massless.
By construction, these amplitudes contain only planar Feynman graphs 
and are constructed with colour-stripped Feynman rules.

The polarization vectors for gluons can be expressed as
\begin{equation}
\e^\mu_{+ } = \frac{\AS{q | \g^\mu | g}}{\sqrt{2}\AA{q g}} \, , 
\quad
\e^\mu_{-} = \frac{\AS{g |\g^\mu | q}}{\sqrt{2}\SSS{g q}} \, ,
\label{pol_gluons}
\end{equation}
where $q$ is the auxiliary light-like vector and $g$ stands for the gluon momentum.
In the expressions of the amplitudes, gluons will be denoted by the number of the corresponding particle, 
whereas we will always explicitly distinguish quarks and antiquarks with $q$ and $\qb$ respectively.

Finally, the polarization vectors for the auxiliary photons coming in pairs with the off-shell quarks are
\beq
\e^\mu_{f\,+} = \frac{\AS{q | \g^\mu | f}}{\sqrt{2}\AA{q f}} \, , 
\quad
\e^\mu_{f\,-} = \frac{\AS{f |\g^\mu | q}}{\sqrt{2}\SSS{f q}} \, ,
\label{pol_photons}
\eeq
where $f$ denotes the fermion momentum spinor and $q$ is again the auxiliary vector.

Further information on the spinor helicity formalism can be found
in~\cite{vanHameren:2014iua,vanHameren:2015bba,vanHameren:2016bfc}.

\section{Real emission splitting functions in a set of variables 
more suitable for an evolution equation}
\label{app:results}

In this section we provide the TMD splitting function in terms of a different set
of variables, that was used in ref.~\cite{Hentschinski:2016wya} in the construction of an evolution equation. 
Namely, we will use the real emitted momentum $\pt'=\kt-\qt$ and the outgoing momentum $\qt$, see Fig.~\ref{real_graphs}.

The definition of the kernel corresponding to Eq.~\eqref{eq:KijangDep} in these variables reads
\beq
\hat K_{ij} \left(z, \frac{\kt^2}{\mu^2}, \epsilon, \alpha_s \right)  =
\frac{\alpha_s \cdot z}{2\pi \mu^{2\epsilon} e^{\epsilon\gamma_E}}  
\int \!\frac{d^{2+2\epsilon}\pt'}{\pi^{1+\epsilon} \pt'^2}  \tilde{P}_{ij}^{(0)}(z,\qt,\pt')
\Theta\left(\mu_F^2 - \frac{z\pt'^2 }{1-z} - \qt^2\right) .
\eeq
In the following we list the splitting functions in the new variables.

\subsection{Angular and transverse momentum dependent TMD splitting functions}

%
\bea
\tilde{P}_{qg}^{(0)} (z,\qt,\pt')
&=& 
T_R\, 
\frac{\pt'^2}{(\pt'+\qt)^2}
\bigg[ 1
     + \frac{2(\pt'\cdot\qt)}{z\pt'^2+(1-z)\qt^2}
     + \frac{\pt'^2\qt^2}{(z\pt'^2+(1-z)\qt^2)^2}
\bigg] \, ,
\eea
\bea
\tilde{P}_{gq}^{(0)} (z,\qt,\pt')
&=& 
C_F\, 
\frac{z\pt'^4 + 2\left(z+\frac{1}{z}-2\right)\qt^4 + 2(1-z)\pt'^2\qt^2}{(z\pt'^2+(1-z)\qt^2)^2}
\nn \\
     &&
+ \epsilon\, C_F \frac{z\pt'^4}{(z\pt'^2+(1-z)\qt^2)^2} \, ,
\eea
\bea
\tilde{P}_{qq}^{(0)} (z,\qt,\pt')
&=& 
C_F\, 
\bigg[ \frac{2}{1-z}
     + \frac{2(\pt'\cdot\qt)}{z\pt'^2+(1-z)\qt^2}
     + \frac{(1-z)\pt'^2\qt^2}{(z\pt'^2+(1-z)\qt^2)^2}
\bigg]
\nn \\
     &&
     + \epsilon\, C_F \frac{(1-z)\pt'^2\qt^2}{(z\pt'^2+(1-z)\qt^2)^2} \, ,
\eea
\bea
 \tilde{P}_{gg}^{(0)} (z,\qt,\pt') 
&=& 
C_A\, 
\bigg[
    \frac{\pt'^4\qt^2}{(\pt'+\qt)^2(z\pt'^2+(1-z)\qt^2)^2}
 \notag \\ &&  - \frac{(2z-1)\pt'^2\qt^2 + 2z\qt^2(\pt'\cdot\qt) + 4(\pt'\cdot\qt)^2}{(\pt'+\qt)^2(z\pt'^2+(1-z)\qt^2)}
\nn \\
&&
  - \frac{z^2(1+z^2)\pt'^4 + z(1-z)(3+z^2)\pt'^2\qt^2 + 2(1-z)^2\qt^4}{z(1-z)(z\pt'^2+(1-z)\qt^2)^2}
 \notag \\
&&  - \frac{2(\pt'\cdot\qt)}{(\pt'+\qt)^2}
\bigg]
   - \epsilon\, C_A \frac{z(1-z)\pt'^2(\pt'^2-\qt^2)^2}{(\pt'+\qt)^2(z\pt'^2+(1-z)\qt^2)^2} \, .
\eea
%

\subsection{Angular averaged TMD splitting functions}
The angular averaged splitting functions listed below are for $\epsilon=0$.
\bea
\bar{P}_{qg}^{(0)} 
&=& 
T_R\, 
\bigg[
       \frac{\pt'^2}{z\pt'^2+(1-z)\qt^2}
     - \frac{z(1-z)\pt'^2\left|\pt'^2-\qt^2\right|}{\left(z\pt'^2+(1-z)\qt^2\right)^2}
\bigg] \, ,
\eea
\bea
\bar{P}_{gq}^{(0)} 
&=& 
C_F\, 
\frac{z\pt'^4 + 2(1-z)\pt'^2\qt^2 + 2\left(z+\frac{1}{z}-2\right)\qt^4}
     {\left(z\pt'^2+(1-z)\qt^2\right)^2} \, ,
\eea
\bea
\bar{P}_{qq}^{(0)} 
&=& 
C_F\, 
\bigg[
       \frac{2}{1-z}
     + \frac{(1-z)\pt'^2\qt^2}{\left(z\pt'^2+(1-z)\qt^2\right)^2}
\bigg] \, ,
\eea
\bea
\bar{P}_{gg}^{(0)} 
&=& 
C_A\, 
\bigg[
     - \frac{2}{z(1-z)}
     + \frac{2\pt'^2}{z\pt'^2+(1-z)\qt^2}
     + \frac{\pt'^2}{\left|\pt'^2-\qt^2\right|}
\nn \\
&&
     - \frac{\pt'^4+\pt'^2\qt^2}{\left|\pt'^2-\qt^2\right|\left(z\pt'^2+(1-z)\qt^2\right)}
     + \frac{\pt'^4\qt^2}{\left|\pt'^2-\qt^2\right|\left(z\pt'^2+(1-z)\qt^2\right)^2}
\bigg] \, .
\eea
%



\end{document}